\documentclass[preprint,pteplogo]{ptephy_v2}

\preprintnumber{XXXX-XXXX} 
\usepackage{hyperref}
\usepackage{subfig}

\begin{document}

\title{Characterization of the 20-inch Photomultiplier Tubes for RENE Detector}




\author[1]{Junkyo Oh}

\author[1]{Byeongsu Yang}
\author[1]{Cheong Heo}
\author[1]{Daeun Jung}
\author[1]{Dong Ho Moon}
\author[1]{Eungyu Yun}
\author[1]{Hyun Woo Park}
\author[1]{Jae Sik Lee}
\author[1]{Jisu Park}
\author[1]{Ji Young Choi}
\author[1]{Kyung Kwang Joo}
\author[1]{Ryeong Gyoon Park}
\author[1]{Sang Yong Kim}
\author[1]{Sunkyu Lee}

\author[2]{Insung Yeo}
\author[2]{Myoung Youl Pac}

\author[3,*]{Jee-Seung Jang}

\author[4]{Eun-Joo Kim}

\author[5]{Hyunho Hwang}
\author[5]{Junghwan Goh}
\author[5]{Wonsang Hwang}

\author[6]{Jiwon Ryu}
\author[6]{Jungsic Park}
\author[6]{Kyu Jung Bae}
\author[6]{SeoBeom Hong}

\author[7]{Hyunsoo Kim}

\author[8]{Dojin Kim}
\author[8]{Jonghee Yoo}
\author[8]{Seunghwan Choi}
\author[8]{Wonjun Lee}

\author[9]{Jubin Park}
\author[9]{Myung-Ki Cheoun}

\author[10]{Intae Yu}

\affil[1]{
    Center for Precision Neutrino Research (CPNR), Department of Physics, 
    Chonnam National University, 77 Yongbong-ro, Buk-gu, Gwangju 61186, Republic of Korea
}

\affil[2]{
    Department of Radiology, Dongshin University,
    67 Dongshindae-gil,
    Naju, Jeollanam-do 58245,
    Republic of Korea
}
\affil[3]{
    Department of Physics and Photon Science, Gwangju Institute of Science and Technology,
    123 Cheomdangwagi-ro, Buk-gu,
    Gwangju 61005,
    Republic of Korea,
}
\affil[4]{
    Division of Science Education, Jeonbuk National University,
    567 Baekje-daero, Deokjin-gu,
    Jeonju, Jeollabuk-do 54896,
    Republic of Korea,
}
\affil[5]{
    Department of Physics, Kyung Hee University,
    26 Kyungheedae-ro, Dongdaemun-gu,
    Seoul 02447, Republic of Korea
}

\affil[6]{
    Department of Physics, Kyungpook National University,
    80 Daehak-ro, Buk-gu,
    Daegu 41566, Republic of Korea
}

\affil[7]{
    Department of Physics and Astronomy, Sejong University,
    209 Neungdong-ro, Gwangjin-gu,
    Seoul 05006, Republic of Korea
}

\affil[8]{
    Department of Physics and Astronomy, Seoul National University,
    1 Gwanak-ro, Gwanak-gu,
    Seoul 08826, Republic of Korea
}

\affil[9]{
    Origin of Matter and Evolution of Galaxy (OMEG) Institute, Department of Physics, Soongsil University, 369 Sangdo-ro, Dongjak-gu,  Seoul, 06978, Republic of Korea
}

\affil[10]{
    Department of Physics, Sungkyunkwan University, Seobu-ro, Jangan-gu, Suwon-si, Gyeonggi-do, 31206, Republic of Korea
}




\begin{abstract}%
To address the Reactor Antineutrino Anomaly (RAA) observed in neutrino experiments, the Reactor Experiment for Neutrino and Exotics (RENE) has been initiated using a liquid scintillation detector. In this study, we investigate the characteristics of two 20-inch Hamamatsu R12860 photomultiplier tubes (PMTs) intended for installation in the RENE detector. The charge and timing responses of the PMTs were evaluated at both the nominal and target gains expected during actual operation. In particular, gain non-uniformity arising from the large-diameter photocathode with a box-and-line type dynode structure was examined, and the maximum gain variation was measured. The occurrence rate, timing, and charge distributions of late pulses and afterpulses were also investigated to characterize the specific response features of the R12860 PMT. The results reported in this study will aid in the interpretation of signals from the RENE detector and serve as a reference for estimating potential systematic uncertainties in RENE data. Furthermore, these findings are expected to provide valuable information for other experiments employing the same type of PMTs.
\end{abstract}

\subjectindex{H20}

\maketitle

\section{Introduction}
The reactor-based experiments on neutrino oscillations, Double Chooz~\cite{DC}, Daya Bay~\cite{DB} and RENO~\cite{RENO}, had reported a consistent result of the final mixing angle $\theta_{13}$ for three-flavor model of neutrino oscillations and contributed to current understanding of neutrino mixing. Despite these advances, however, the Reactor Antineutrino Anomaly (RAA) remains unsolved. Physicists speculate that this anomaly is attributed either to the existence of a sterile neutrino or to currently unidentified components of reactor. Building on these findings, a joint analysis conducted by the NEOS~\cite{NEOS} and RENO~\cite{RENO_sterile} collaboration has identified a tightly constrained region for the oscillation parameters associated with sterile neutrinos~\cite{joint_RENO_NEOS}. Moreover, to further investigate the reactor antineutrino spectrum, several short-baseline reactor experiments are currently underway~\cite{DANSS, PROSPECT, STEREO}. In this context, the newly proposed Reactor Experiment for Neutrino and Exotics (RENE)~\cite{RENE} aims to probe the allowed sterile neutrino parameter space suggested by the joint analysis of NEOS and RENO, and to obtain a more detailed reactor neutrino spectrum for a better understanding of the RAA. It will be conducted at the same site as NEOS, specifically in the tendon gallery of one of the reactors at the Hanbit Nuclear Power Plant in Yeonggwang, Jeollanam-do, Republic of Korea. The RENE detector is a gadolinium (Gd) loaded liquid scintillator (LS) detector equipped with two 20-inch photomultiplier tubes (PMTs). Before its construction, the basic performance and feasibility of employing these large-diameter PMTs for the RENE experiment were evaluated under \textit{ex-situ} conditions. These \textit{ex-situ} measurements are expected to help interpret \textit{in-situ} results and to be beneficial not only for RENE but also for other various particle physics experiments since this large diameter PMTs are widely employed in experiments.

\section{RENE detector and R12860 PMTs}
\begin{figure}[hbt!]
\centering
\includegraphics[width=0.8\textwidth]{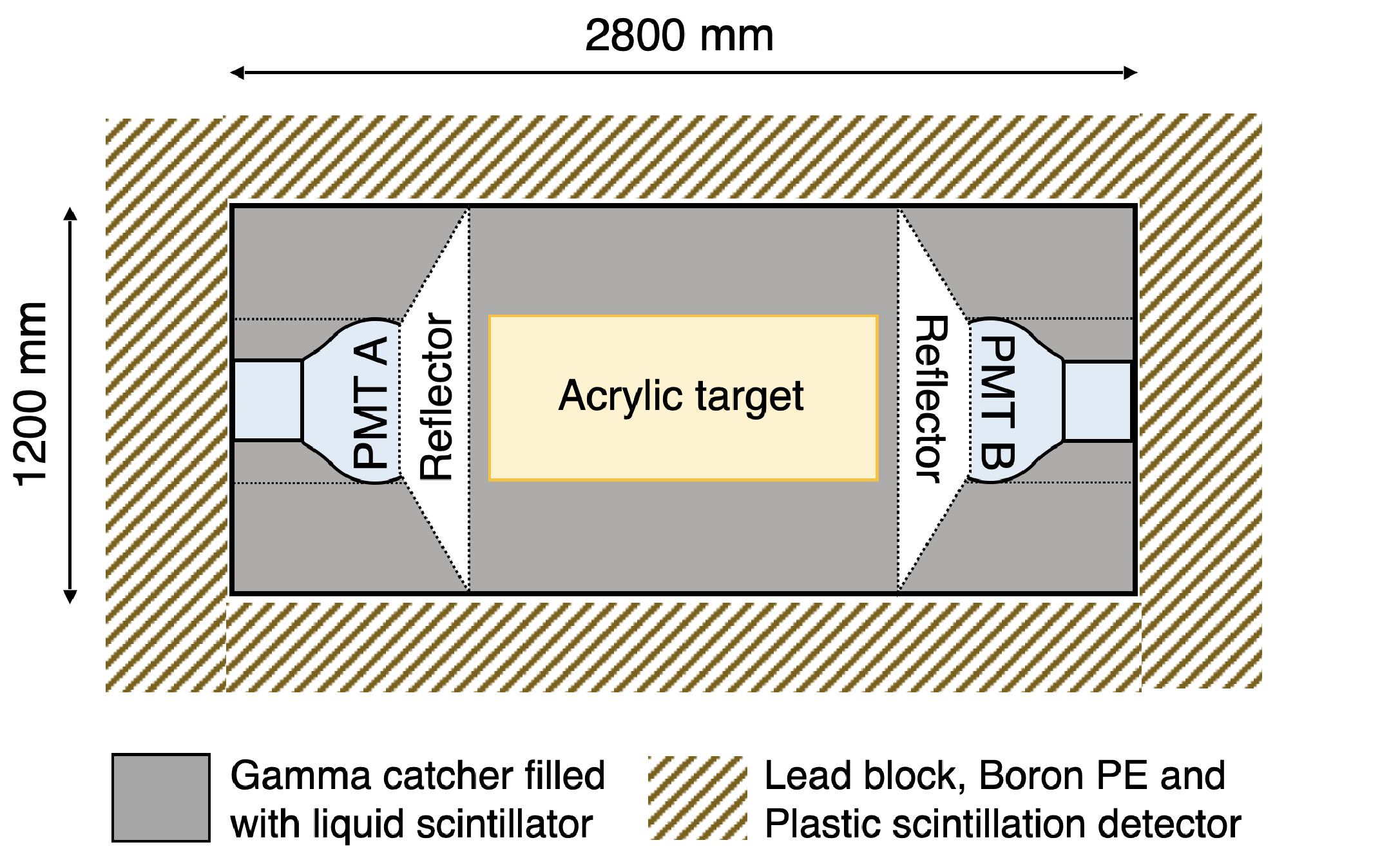}
\caption{\setlength{\baselineskip}{4mm} 
    Schematic drawing of the RENE detector (top view). PMTs are installed inside the gamma-catcher (width:depth:height = 2800:1200:1200 mm).}
\label{fig:detector_rene_geometry}
\end{figure}

\begin{figure} [!htb]
\centering
        \includegraphics[width=0.9\textwidth]{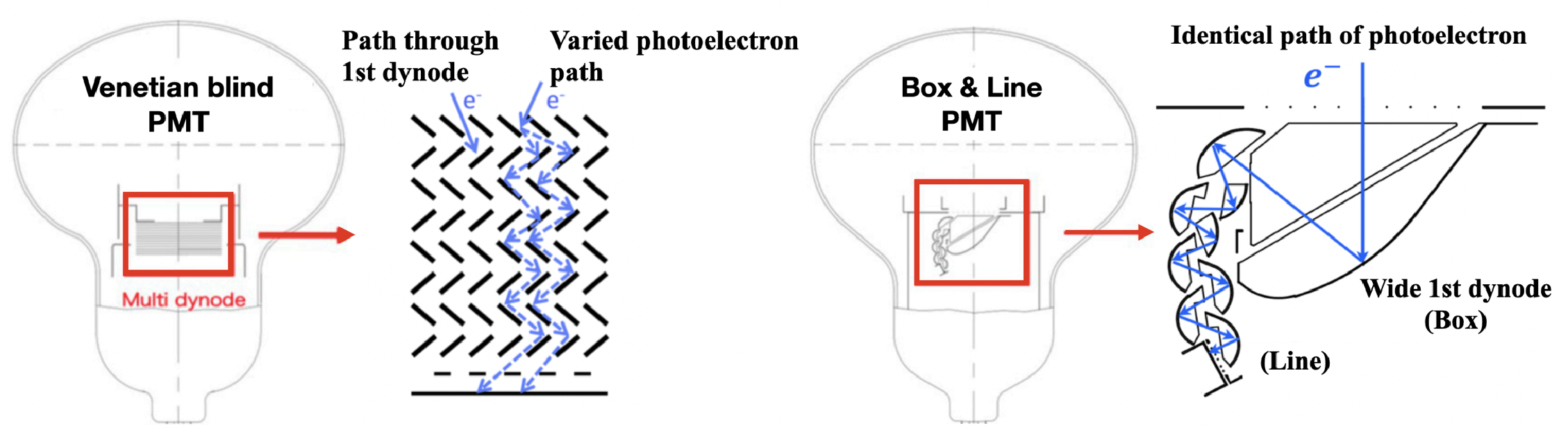}
    \caption{Structure of venetian blind dynode system in R3600 (left) and box-and-line dynode system in R12860 (right)~\cite{dynode, semuli}.}
    \label{fig:dynode}
\end{figure}

Fig.~\ref{fig:detector_rene_geometry} presents a schematic of the RENE detector. This detector is composed of a target vessel filled with the Gd-LS (270 L), a gamma-catcher chamber filled with the LS ($\sim$3000 L) and PMTs enclosed by a reflector cone. The PMTs are installed facing each other, and the chamber is surrounded by layers of Boron polyethylene (PE) and lead blocks. A veto detector system incorporating plastic scintillation detectors is installed in the outermost layer. Notably, the RENE adopts Hamamatsu R12860 PMTs, oil-proof 20-inch PMT developed with a modified dynode structure to improve the collection efficiency of photoelectrons. These PMTs achieve better performance than the R3600 PMTs used in Super-Kamiokande~\cite{SK} detector, as illustrated in Fig.~\ref{fig:dynode}~\cite{dynode}. The specifications of R12860 PMTs used in the RENE detector are provided in Tab.~\ref{tab:R12860}.

\begin{table}[htbp]
\centering

\begin{tabular}{l|c}
\hline
Parameter & Description and values \\
\hline
Spectral response & 300 nm to 650 nm\\
Peak wavelength & 420 nm\\
Photocathode material & Bialkali \\
Dynode structure & Box and Line\\
Quantum efficiency at 390 nm & $30\%$ \\
Nominal gain & $1.0\times10^{7}$ \\
Applied voltage for nominal gain & 2000 V\\
\hline
\end{tabular}
\caption{The specifications of R12860 PMTs~\cite{R12860}.\label{tab:R12860}}

\end{table}

\section{Experimental setup for \textit{ex-situ} measurements}

\begin{figure} [!htb]
\centering
\includegraphics[width=1.0 \textwidth]{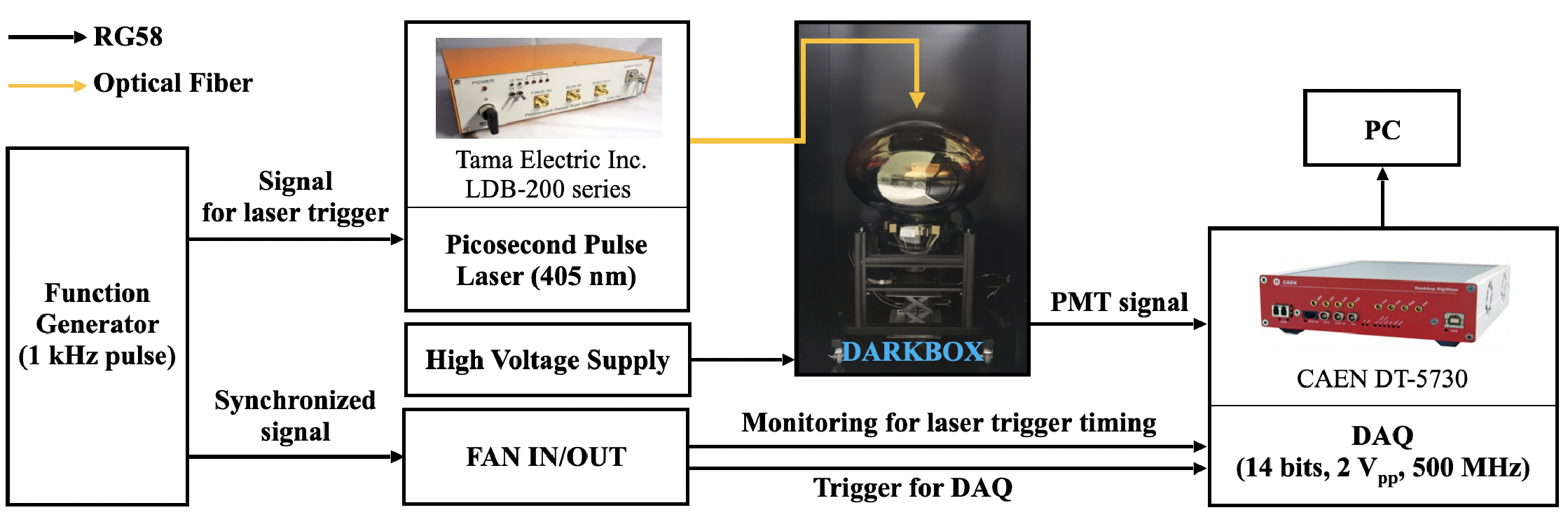}
    \caption{Schematic diagram for the test setup.}
    \label{fig:DAQsetup}
\end{figure}

To investigate the fundamental characteristics of PMT, all light other than the measurement source was carefully blocked. The PMT, placed inside a dark box, was passively shielded from the geomagnetic field with a mu-metal enclosure and the residual magnetic field was maintained at less than 100 mG. Even at magnetic-field levels below 100 mG, measurable deviations in both the PMT gain and quantum efficiency can still occur, as the residual magnetic field perturbs the electron trajectories within the dynode structure. However, the PMTs installed in the RENE detector are passively shielded with mu-metal, which reduces the residual magnetic field to below 100 mG. To characterize the PMT performance under conditions relevant to the RENE experiment, all measurements were conducted while allowing for the level of uncertainty associated with a comparable residual magnetic field. In addition, the temperature and humidity at the RENE experimental site are expected to be similar to those of the NEOS experiment~\cite{Neutrino2020_neos}, since RENE is planned to be conducted in the tendon gallery as the NEOS. During this measurement, the temperature and humidity were maintained at $22\pm2^{\circ}\mathrm{C}$ and $60\pm5\%$, respectively, to match the anticipated environmental conditions of the RENE site. The measurement light was delivered through an optical fiber and illuminated the PMT cathode surface within approximately 3 cm. A 405 nm laser with a picosecond pulse width, operated at a repetition rate of 1 kHz, was employed as the light source. To maximize the signal-to-noise ratio, the laser trigger was simultaneously used as the trigger for Data Acquisition (DAQ) as shown in Fig.~\ref{fig:DAQsetup}. Furthermore, the laser trigger signal was monitored by the DAQ to measure the PMT response time. A DAQ system with 14-bit resolution and a sampling rate of 500 MHz was employed in this measurement.

\section{Measurements}
\subsection{Charge and timing response to single-photoelectron (SPE)}
\begin{figure} [!htb]
\centering
\includegraphics[width=0.9\textwidth]{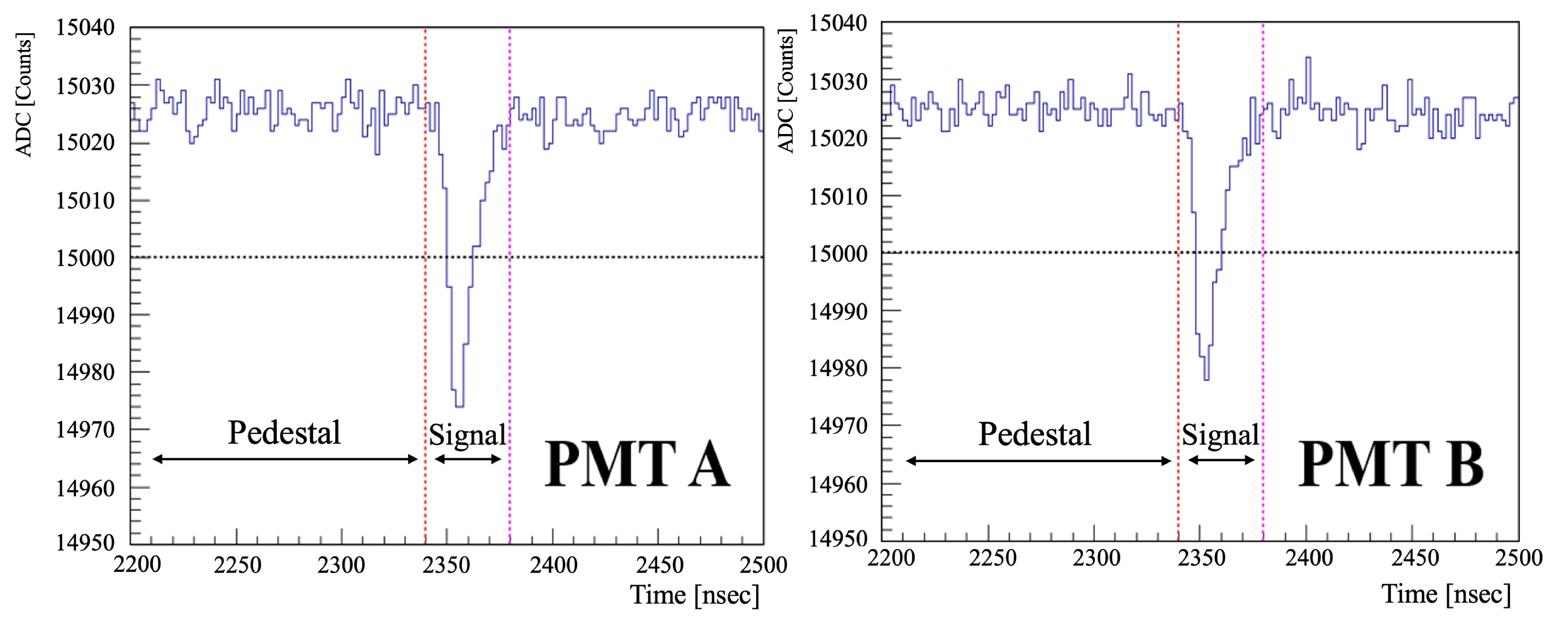}
    \caption{Typical pulse shapes for the two PMTs in response to picosecond laser light~\cite{RENE}. The vertical axes show the analog-to-digital converter (ADC) counts and the dotted horizontal line indicates the 3 mV threshold used for signal selection corresponding to approximately 0.3 p.e.}
    \label{fig:pulse}
\end{figure}

\begin{figure} [!htb]
\centering
\includegraphics[width=1.0 \textwidth]{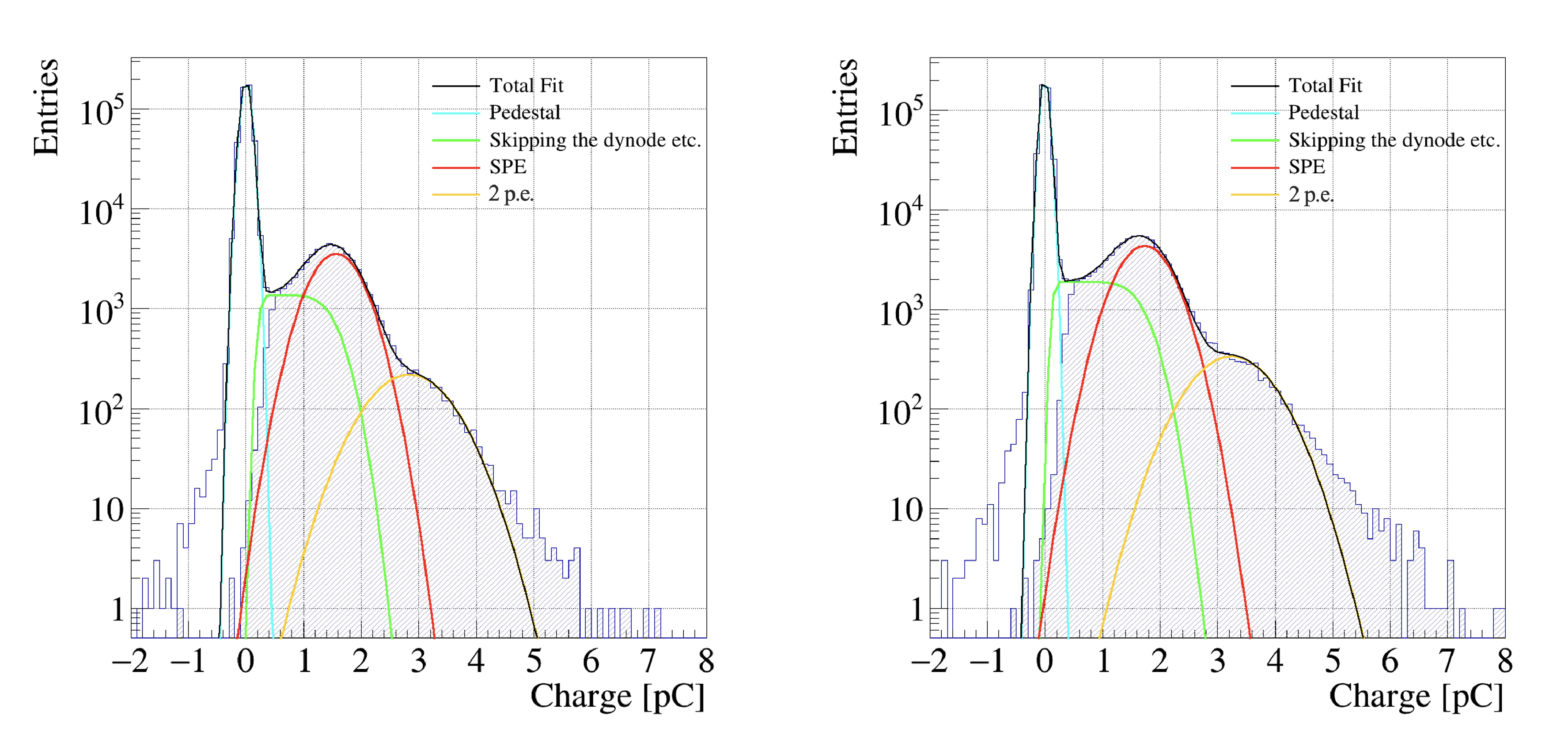}
    \caption{Charge distributions of the PMT signals. The pedestal, SPE and 2 p.e. are fitted using Gaussian function. An error function is used to fit the residual charge between the pedestal and SPE peak.}
    \label{fig:qdist}
\end{figure}

Fig.~\ref{fig:pulse} presents typical pulse shapes of the two PMTs in response to light emitted from picosecond pulse laser. Most of the signals induced by laser appear within the defined signal region, as the synchronized signal of laser trigger is used to active DAQ. Meanwhile, the pedestal for each DAQ channel is calculated using the average value within a designated pedestal calculation window. To achieve a single-photon regime, the intensity of the incident light is sufficiently attenuated to ensure that signals exceeding the 0.3 photoelectrons (p.e.) threshold are detected in fewer than $10\%$ of the injections. Fig.~\ref{fig:qdist} displays the corresponding charge distributions for the two PMTs under single-photon illumination. The pedestal, SPE, and 2 p.e. distributions were determined by fitting the charge distribution with Gaussian and error functions. The flat components between the pedestal and SPE distribution, modeled by an error function, can arise from processes such as photons passing through the photocathode and directly striking the first dynode, or photoelectrons bypassing some dynodes due to back-scattering from the dynode~\cite{erf}. The charge resolutions of the SPE distributions were measured to be $26.0\%$ and $24.1\%$ with peak-to-valley ratios of 3.0 and 2.9 for PMT A and B, respectively. 

\begin{figure}[!htb]
\centering
\includegraphics[width=0.9\textwidth]{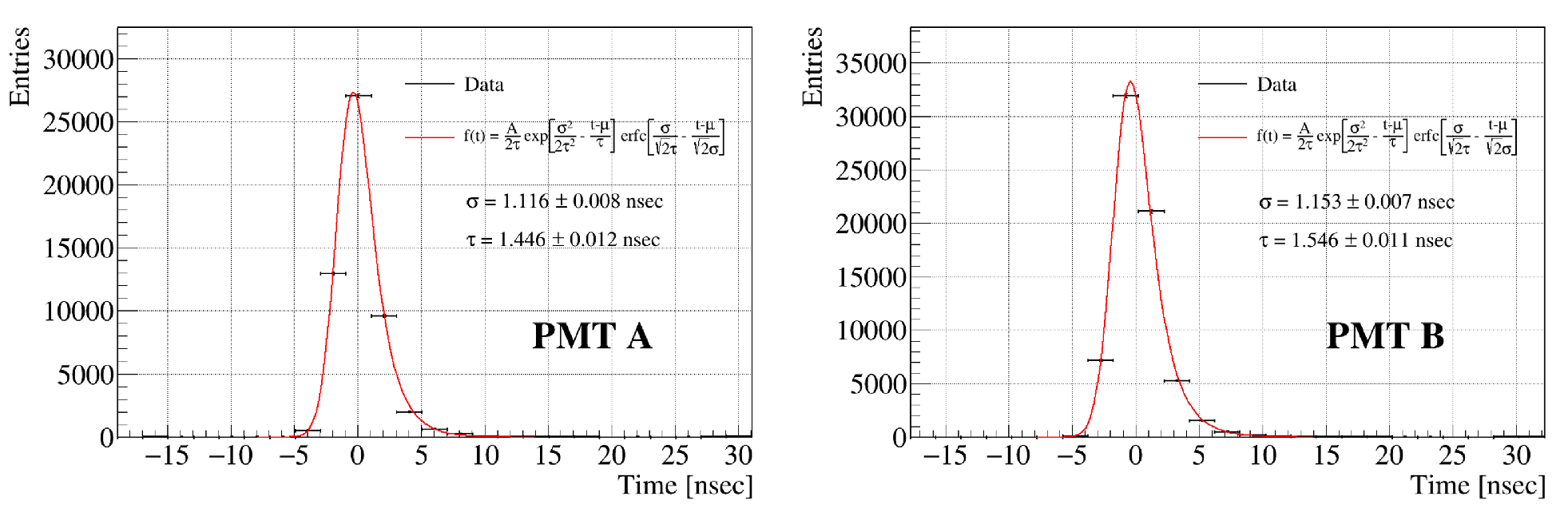}
    \caption[TTS measured using SPE signals]{TTS measured using SPE signals at $1 \times 10^7$ gain. The red trend line represents the convolution of a Gaussian and an exponential function: 
    $f(t) = \frac{A}{2\tau} \exp\left( \frac{\sigma^2}{2\tau^2} - \frac{t - \mu}{\tau} \right) \mathrm{erfc}\left( \frac{\sigma}{\sqrt{2}\tau} - \frac{t - \mu}{\sqrt{2}\sigma} \right)$.}
    \label{fig:tdist}
\end{figure}

The response time of the PMTs was measured using the laser trigger time as a timing reference. The hit timing was defined as the point at which the falling edge of the PMT pulse crossed a threshold corresponding to approximately 0.3 p.e. Fig.~\ref{fig:tdist} shows the difference between the timing of the laser injection trigger and the hit timing of each PMT, which reflects the transit time spread (TTS) of photoelectrons in the PMTs. The TTS for both PMTs was measured to be around 3.5 nsec with $1 \times 10^7$ gain. The FWHM (full width at half maximum) of the convolution of a Gaussian function and an exponential function was taken as the TTS. Restricting the Gaussian fit to the central peak of the distribution yields a TTS of approximately 2.6 nsec for both PMTs, in good agreement with the manufacturer’s specification~\cite{R12860}. The laser response delay and the unexpected trajectories of photoelectrons contribute to the exponential component of the TTS. A conservative TTS was quoted in this study, including all uncertainties arising from these effects and the both PMT exhibited comparable TTS values.

\subsection{Gain measurement}

\subsubsection{Gain stability}
\begin{figure} [!htb]
\centering
\includegraphics[width=0.9\textwidth]{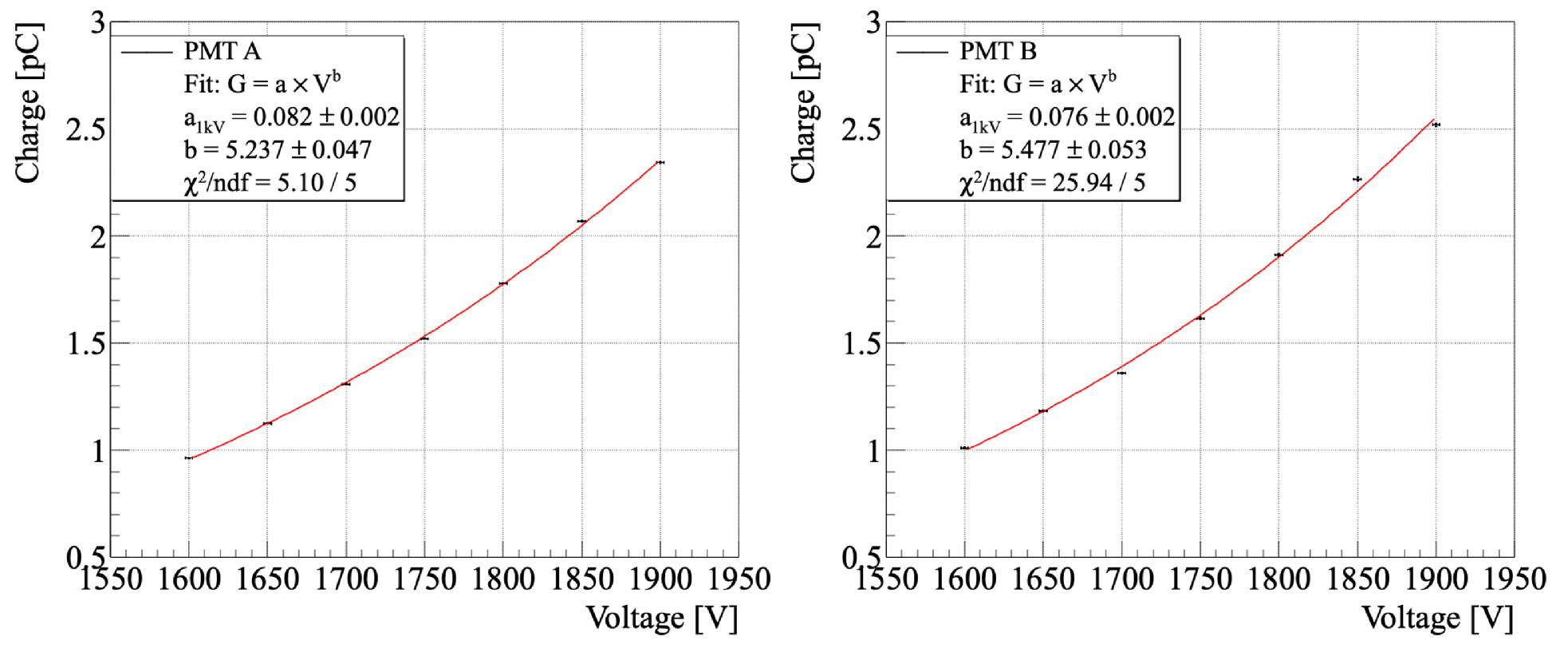}
    \caption{Gain as a function of applied voltage. The results follow the power-law relationship $G(V)=\alpha \times V^{\beta}$, G is the charge corresponding to the PMT gain, and V is the applied voltage.}
    \label{fig:gain}
\end{figure}

\begin{figure} [!htb]
\centering
\includegraphics[width=0.6\textwidth]{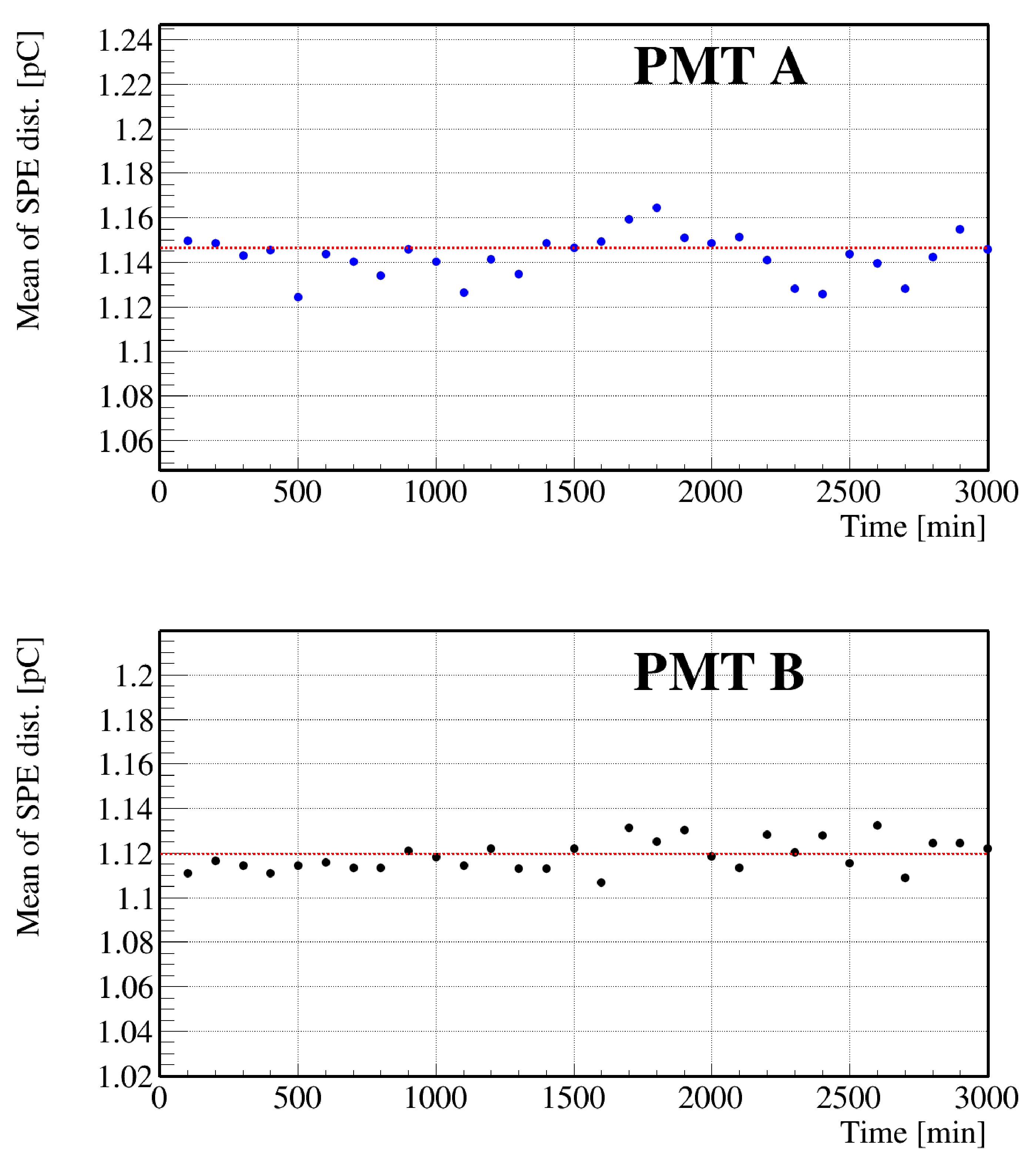}
    \caption{Stability of the mean SPE charge at 1680 V for PMT A and 1650 V for PMT B.}
    \label{fig:stability of gain}
\end{figure}

The nominal gain of the R12860 PMTs is the $1 \times 10^7$ at an applied voltage of 2000 V. Fig.~\ref{fig:gain} displays the gain as a function of the applied voltage. As expected, the gain follows a power-law relationship, $G(V)=\alpha \times V^{\beta}$. Here, $\alpha$ reflects intrinsic properties determined by the PMT's geometric configuration and materials, serving as a proportional constant that sets the overall gain. Meanwhile, the exponent $\beta$, which depends on the number of dynode stages, characterizes how sensitively the gain responds to changes in voltage. 
The gain was measured by exposing the center of the PMT cathode to the single-photon illumination. Since the gain stability of the PMT directly affects the detector's energy resolution, it should be guaranteed. Thus, during operation of detector, the gain will be monitored and calibrated using either radioactive sources or a light source. Before the PMTs are installed in the RENE detector, their gain stability was evaluated with single-photon illumination. During this evaluation, the gain was set to approximately $7\times10^6$, which corresponds to the operational value adopted in the RENE detector to minimize the effects of the PMT saturation~\cite{SK}. Fig.~\ref{fig:stability of gain} shows the variation of the mean SPE charge over 3000-minutes periods, indicating that both PMTs maintained their gain within $\pm 2\%$, although PMT A exhibited slightly larger fluctuations.

\subsubsection{Position dependence of the gain}

\begin{figure} [!htb]
\centering
\hspace{0.5cm}
\includegraphics[width=0.8\textwidth]{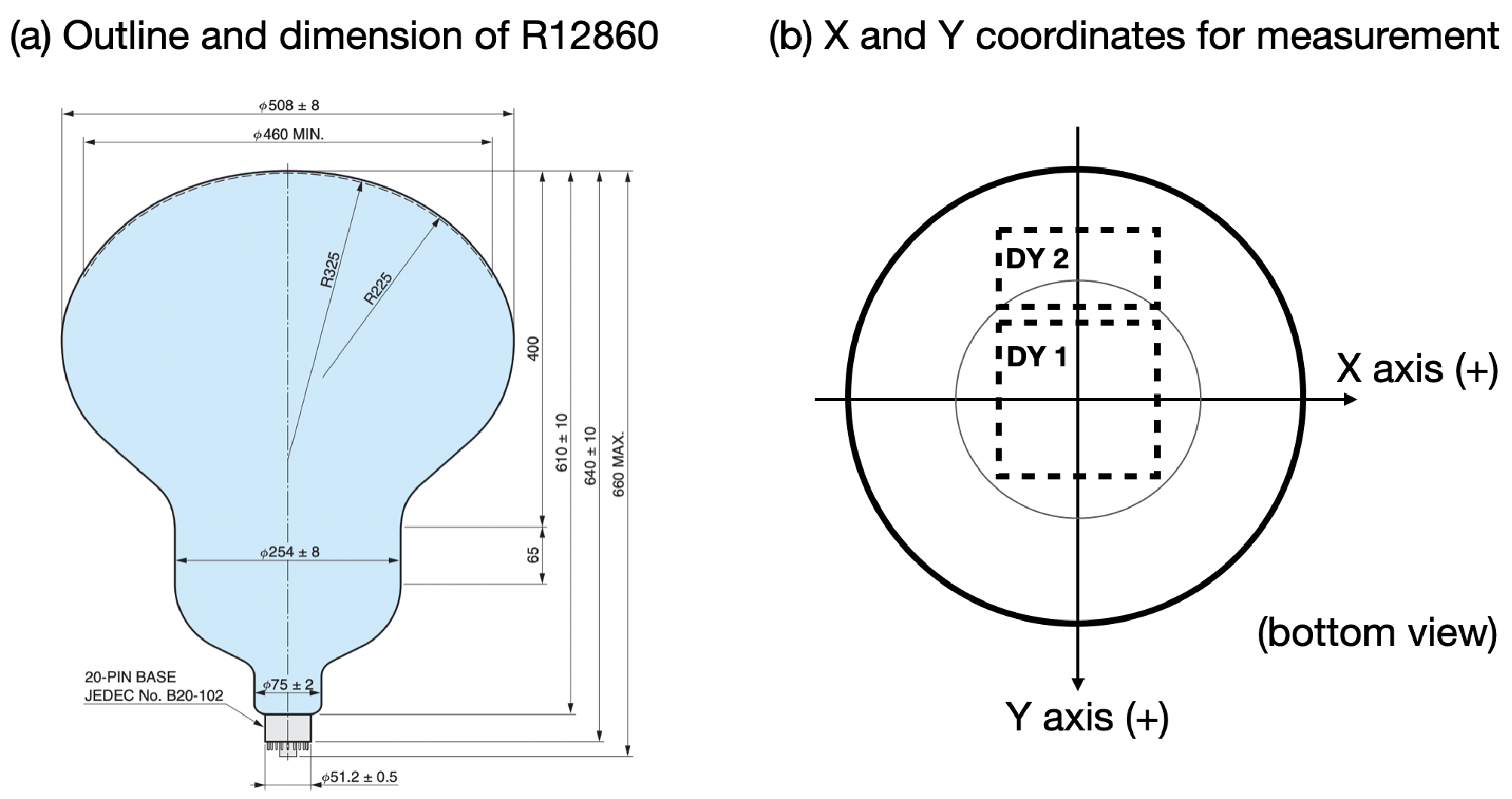}
\caption{(a) Outline and dimensions of the R12860 PMTs~\cite{R12860}. (b) X and Y coordinates viewed from the bottom of the PMT. DY1 is the wide first dynode, and DY2 is the linearly aligned dynodes shown in Fig.~\ref{fig:dynode}.}
\label{fig:PMT_structure}
\end{figure}

\begin{figure} [!htb]
\centering
\includegraphics[width=0.9\textwidth]{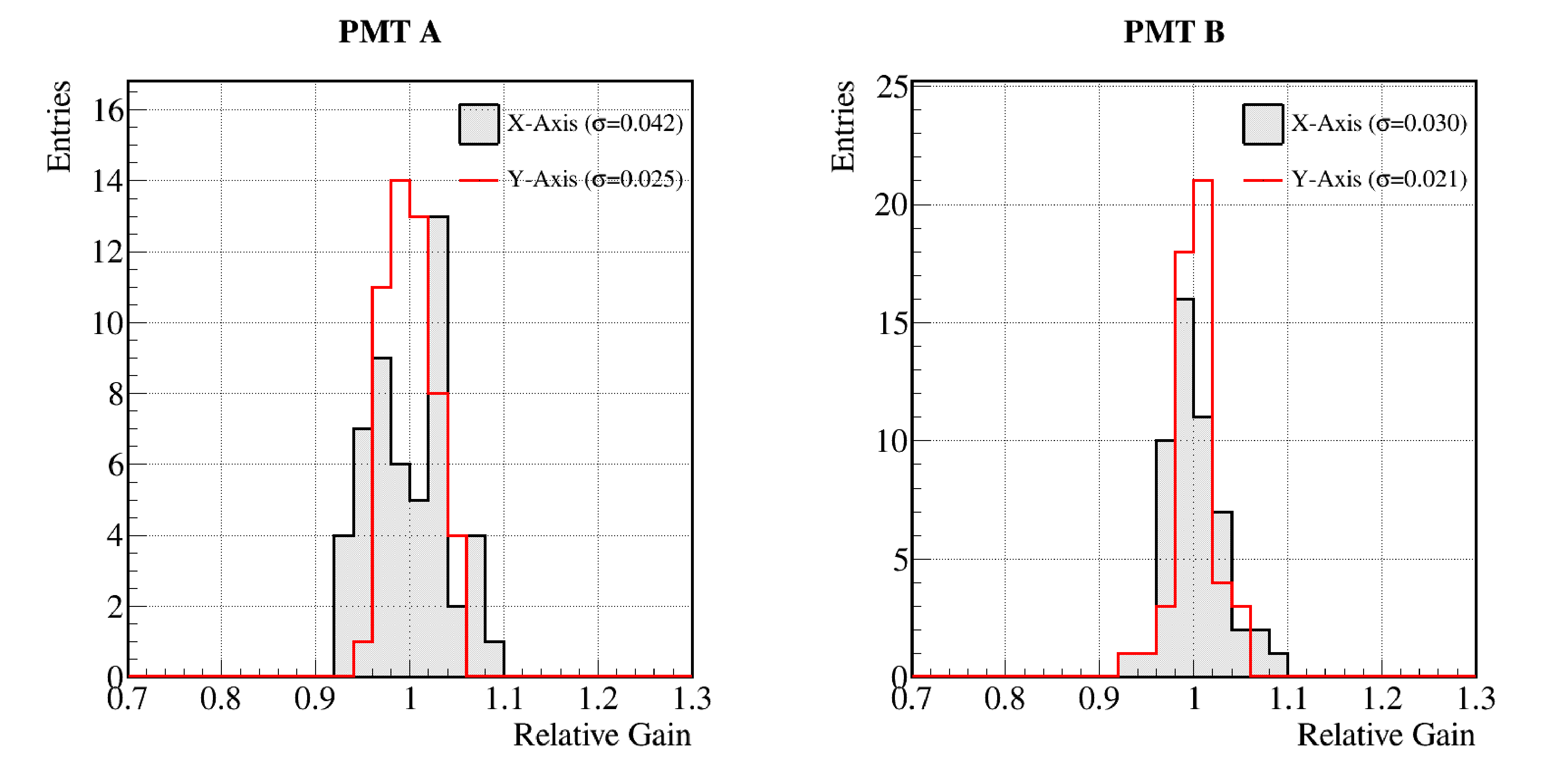}
    \caption{The relative gain for each PMT. Measured values are normalized to the average of all measurements for each axis.}
    \label{fig:relative gain}
\end{figure}

The variation in gain with temperature or magnetic field is well known~\cite{hamamatsu, magnetic field effect}. During the operation of detector, the temperature will be kept constant using air conditioner, and the geomagnetic field will be compensated by mu-metal. However, gain variation along the photocathode surface cannot be controlled and it may degrade the energy resolution. In the RENE detector, a cone-shaped reflector surrounding PMT's cathode (Fig.~\ref{fig:detector_rene_geometry}) collects the scintillation light and the entire area of 20-inch cathode will be utilized for photon detection. Gain variation with respect to the interaction position on the cathode was measured along the X- and Y-axes for the two 20-inch PMTs, as illustrated in Fig.~\ref{fig:PMT_structure}. When viewed from the bottom of the PMT, the X-axis is defined as the direction passing through DY1 only, along which the dynodes at the +X and -X positions are symmetrically arranged. The Y-axis is defined as the direction passing through both DY1 and DY2, along which the dynodes are more widely distributed than those along the X-axis. Since these two axes are expected to produce the most extreme variations based on the dynode geometry, measurements were performed along both the X and Y directions. The gain along the X-axis varied by up to $\pm10\%$ for both PMTs, and the magnitude of gain variation along the Y-axis was smaller than that observed along the X-axis as shown in Fig.~\ref{fig:relative gain}. This behavior was reproducibly observed for both PMTs to be installed in the RENE detector and this weakly suggests that the position-dependent gain is influenced by the dynode structure. Our measurements were performed at intervals of approximately 30--50 mm along each axis, covering a range of up to $\pm200$ mm. Because the measurement positions were manually set, the relative positions along each axis carried comparatively larger uncertainties in their absolute locations. In contrast, the X- and Y-axes used for the measurements could be well defined based on the dynode orientation and the direction of the cable outlet provided by the manufacturer. The development of a system that guarantees precise positioning would enable a more detailed characterization of the gain variation with respect to the measurement location. However, given the positional uncertainties in the current measurements, this study focused on presenting the rate of gain variation along each axis. 
Experiments employing large-diameter PMTs have recently been actively investigating position dependence in order to improve the accuracy of energy and momentum-direction measurements of detected particles~\cite{uniformity}.

\subsection{Timing response characteristics}

\subsubsection{Transit time spread (TTS)}

\begin{figure} [!htb]
\centering
\includegraphics[width=0.9\textwidth]{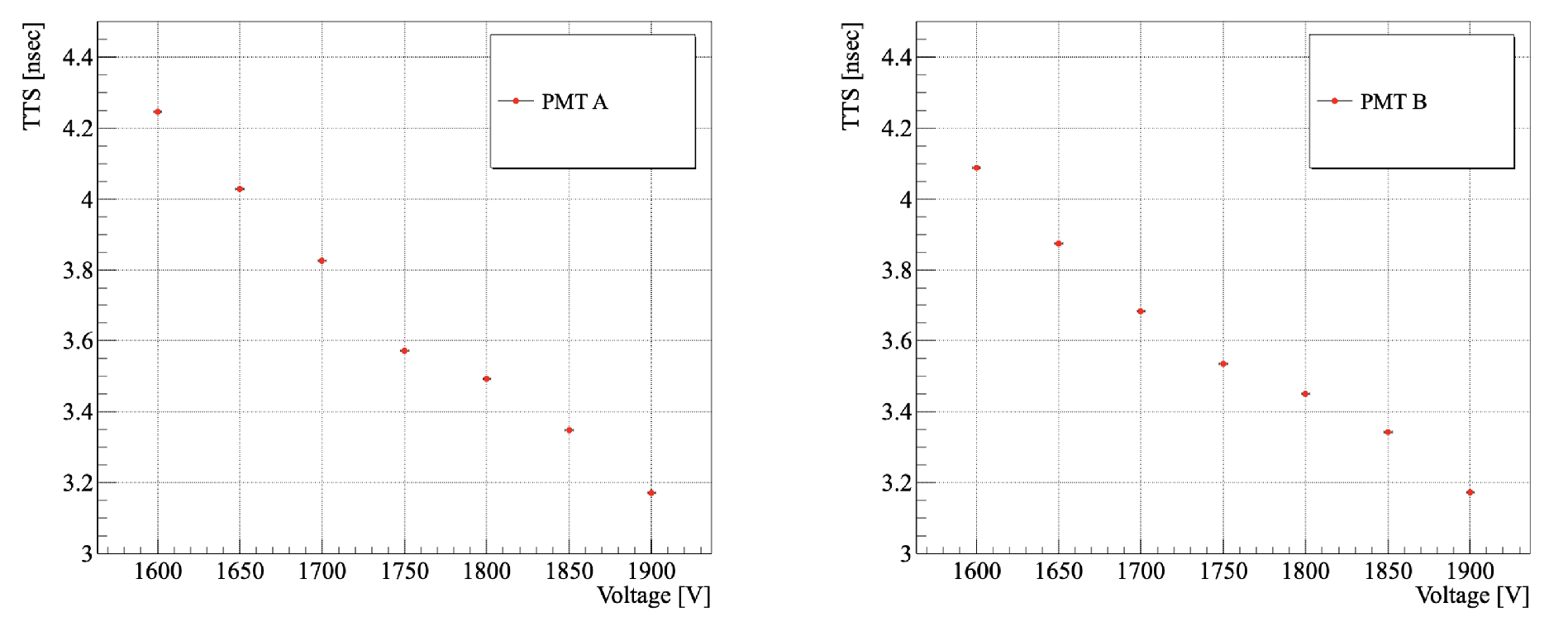}
    \caption{TTS as a function of the applied voltage.}
    \label{fig:TTS}
\end{figure}

Since the RENE detector will be operated with several different gain considering the experiment condition, the timing response was measured as function of the applied voltage. The results follow the well-known behavior of TTS~\cite{hamamatsu} as shown in Fig.~\ref{fig:TTS}, and the TTS at the RENE target gain of $7\times10^{6}$ was measured to be below 4 nsec for both PMTs. While this measurement was performed at the center of the cathode, a position-dependent variation in the TTS is anticipated due to differences in the travel distance for photoelectrons generated at the various positions across the large cathode surface, with the variation anticipated to be a few nanoseconds. Nonetheless, because the RENE experiment detects neutrinos using scintillation light, such variations in PMT response time are considered negligible.

\begin{figure} [!htb]
\centering
\includegraphics[width=0.65\textwidth]{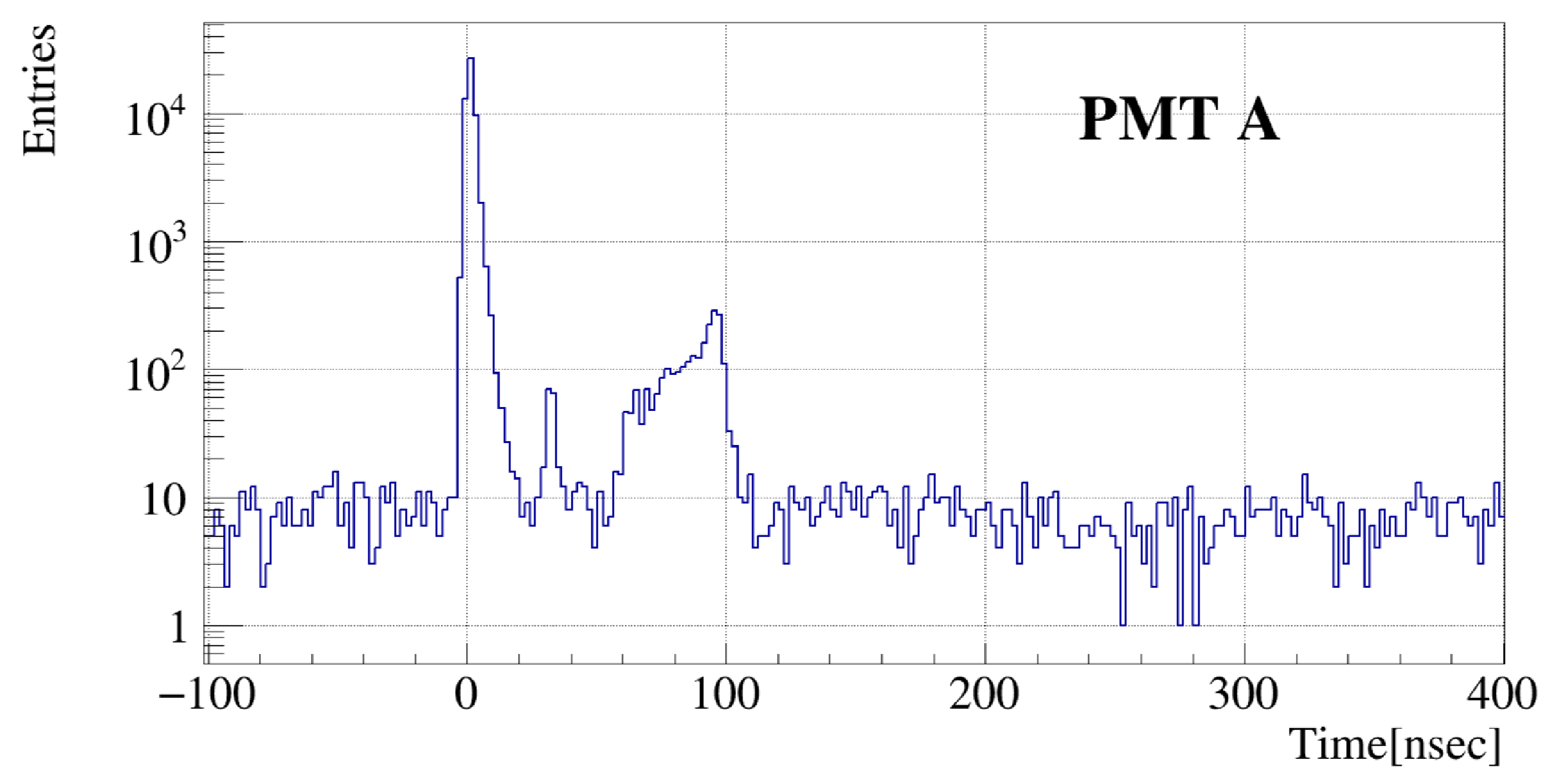}
\includegraphics[width=0.65\textwidth]{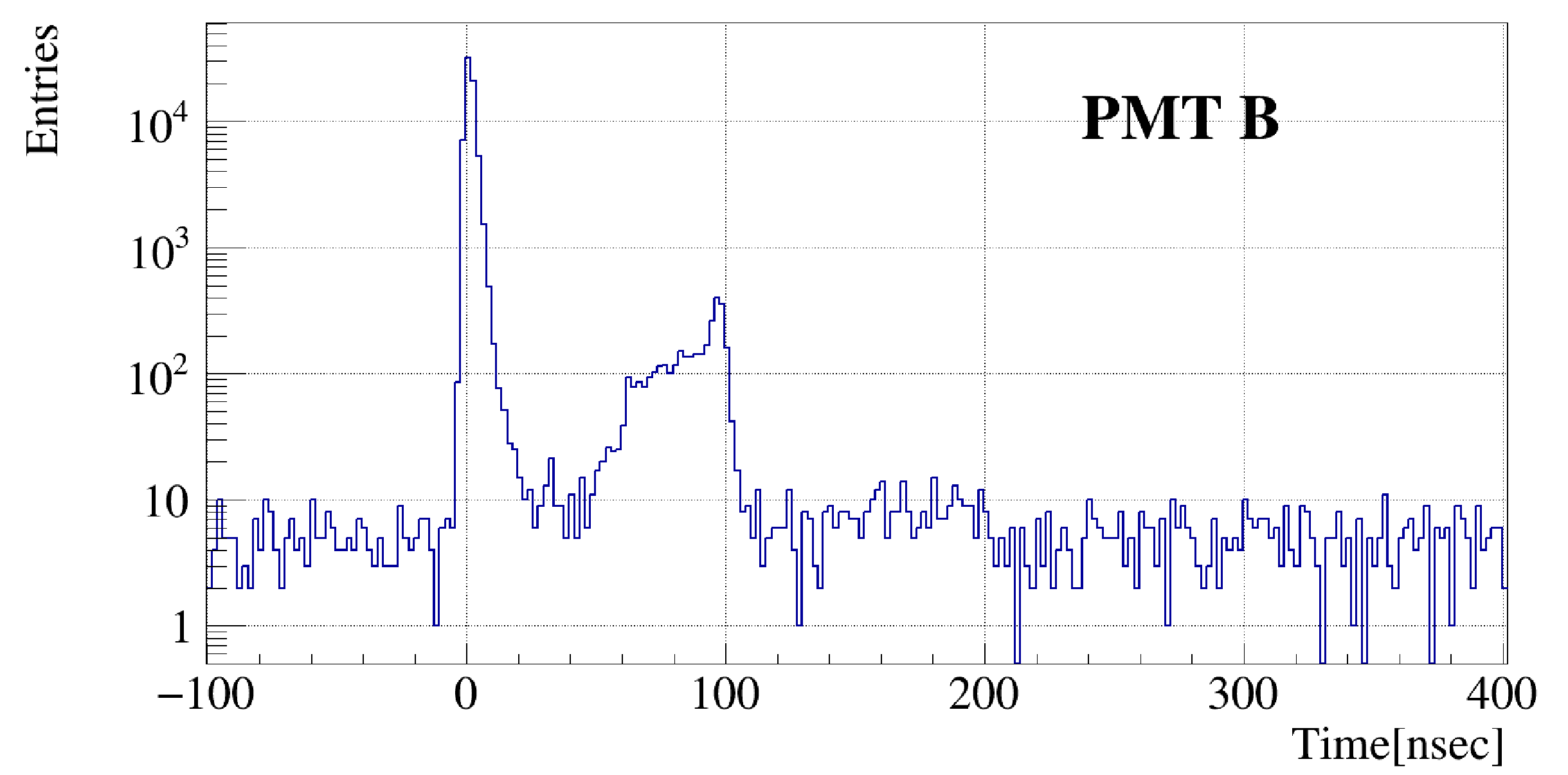}
\caption{Timing distribution for each PMT. Late pulses clearly appear approximately 100 nsec after the main pulse. A smaller peak around 40 nsec arises from reflected light at the fiber connection part.}
\label{fig:latepulse}
\end{figure}

\subsubsection{Late pulses}

For timing response measurement, a picosecond pulse laser was used, allowing detailed observation of the timing response. Fig.~\ref{fig:latepulse} is the difference between the trigger timing for laser injection and hit timing of PMTs. The first peak at 0 nsec corresponds to the timing of the main signal, while a second distinct peak appeares approximately 100 nsec later. This second peak originates from late pulse. This late pulse should be distinguished from afterpulse. Late pulses arise from the elastic or inelastic backscattering of photoelectrons at the first dynode~\cite{latepulse}. These backscattered photoelectrons are initially decelerated by the electric field produced by the applied voltage and then re-accelerated toward the first dynode. The clear late pulse of 20-inch PMT are observed around 100 nsec after main pulse and the occurrence rate of such late pulses is approximately $1\%$ of the main pulses. These late pulses were not included in the estimation of gain presented in Fig.~\ref{fig:qdist}.

\subsubsection{Afterpulses}

\begin{figure}[!]
    \centering
    \subfloat{\includegraphics[width=0.5\textwidth]{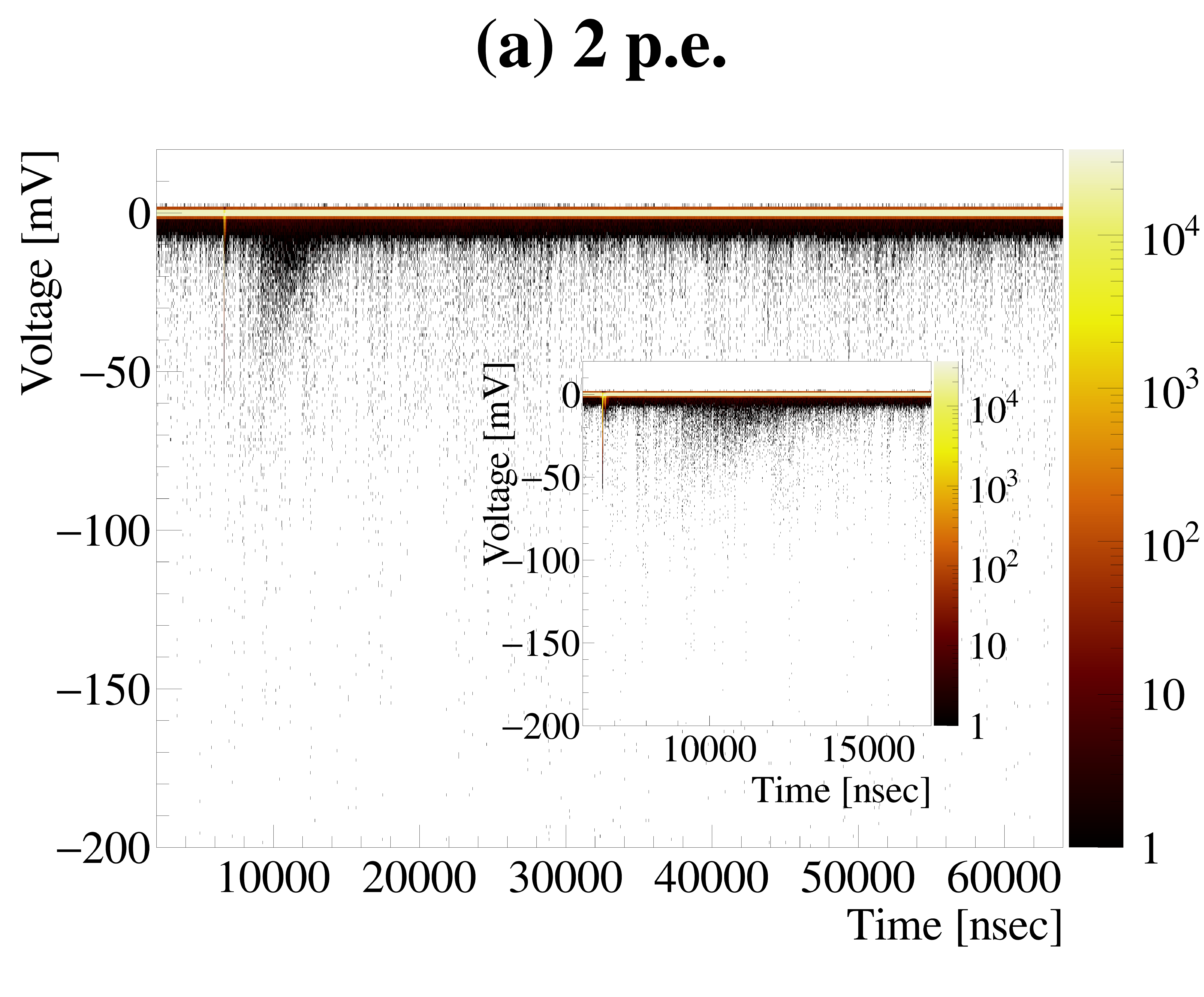}}
    \subfloat{\includegraphics[width=0.5\textwidth]{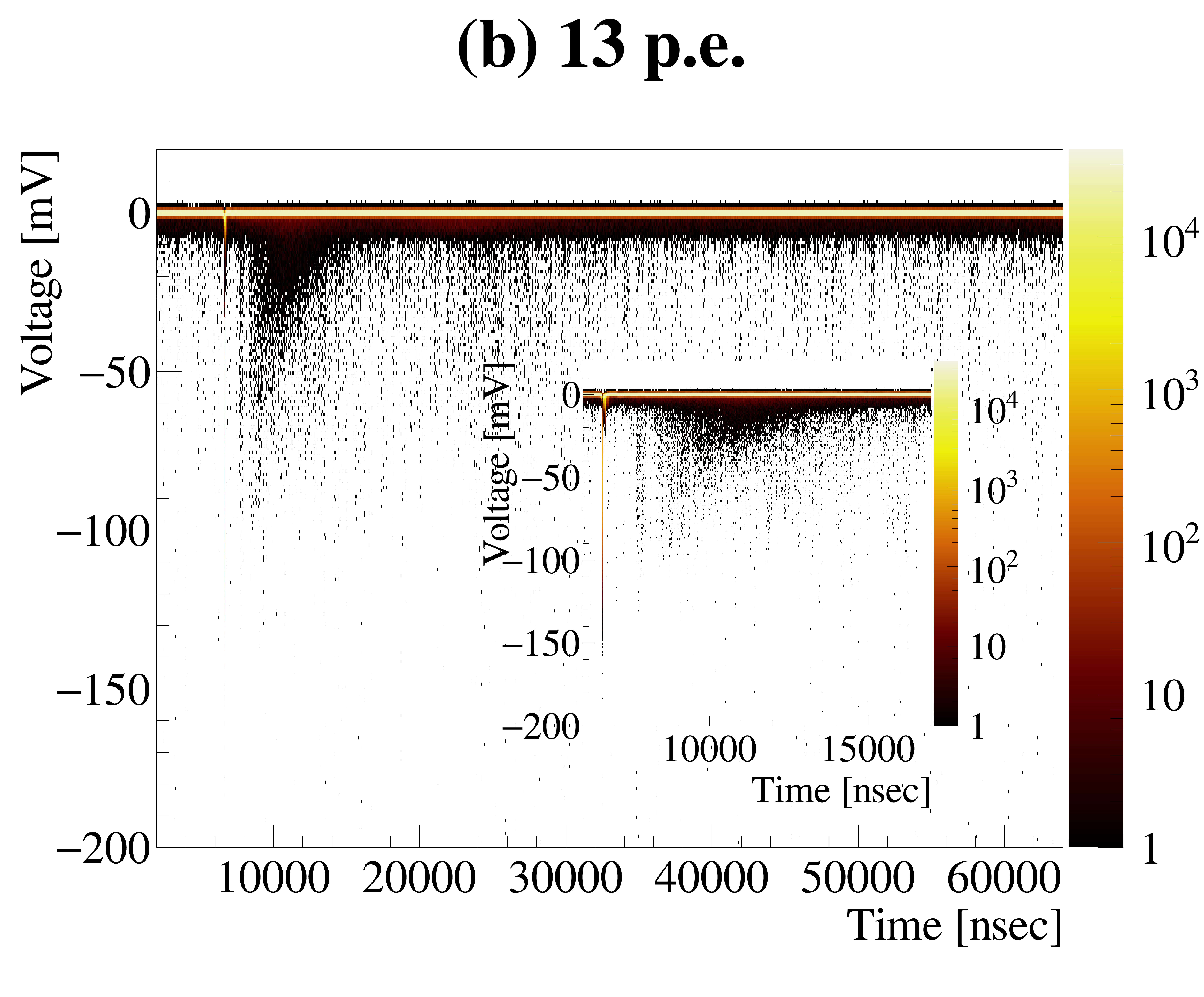}}\\
    \subfloat{\includegraphics[width=0.5\textwidth]{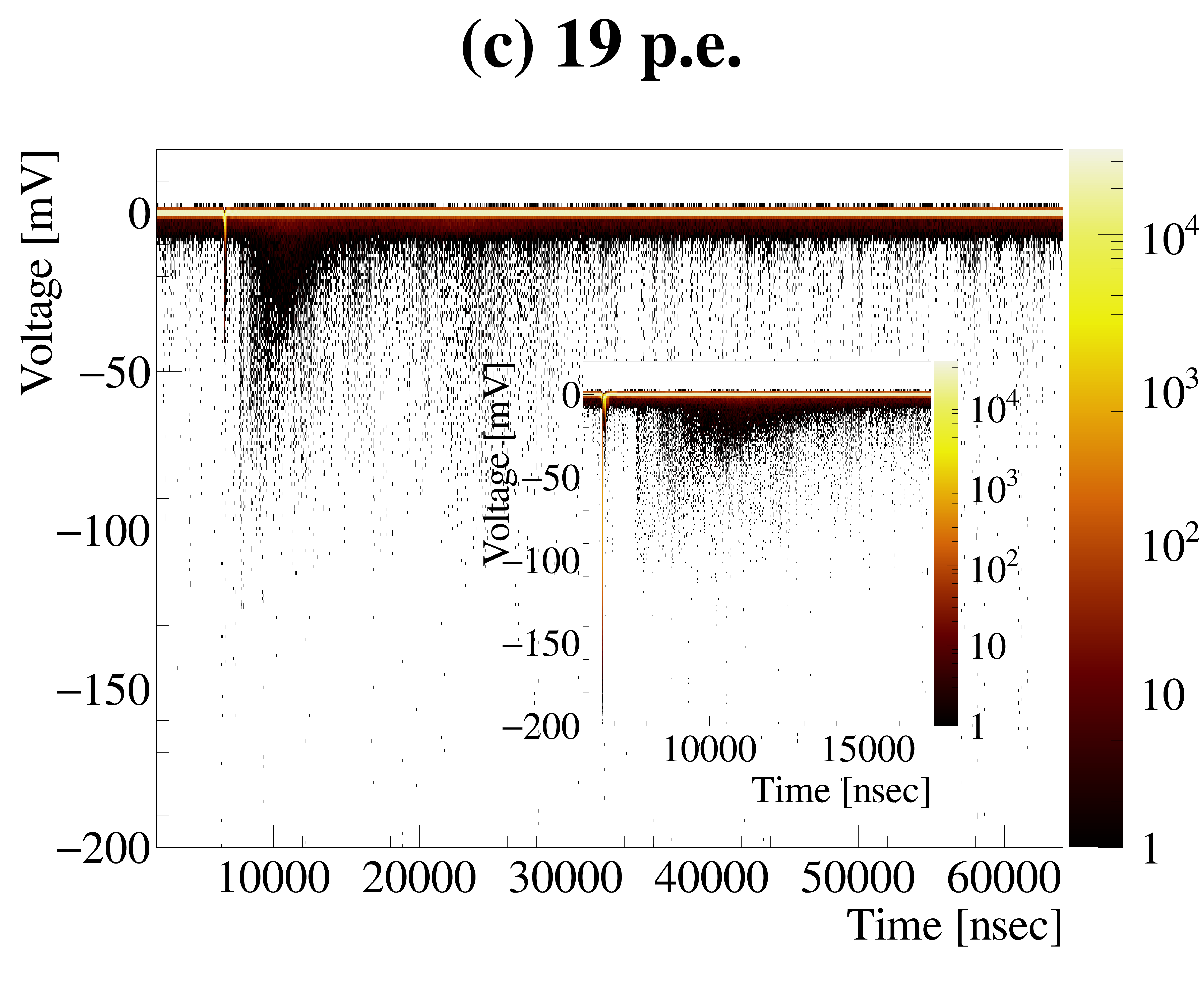}}
    \subfloat{\includegraphics[width=0.5\textwidth]{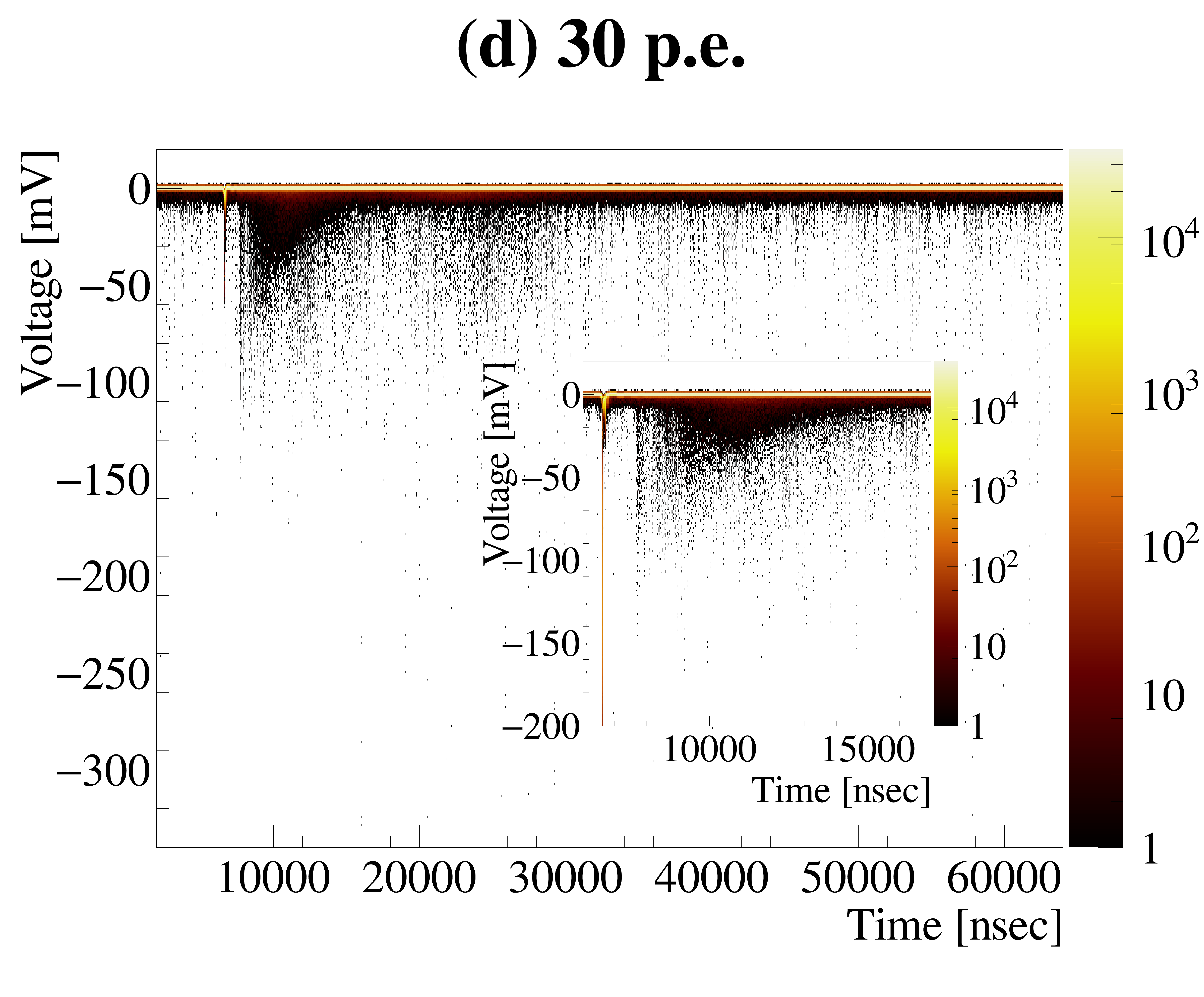}}\\
    \subfloat{\includegraphics[width=0.5\textwidth]{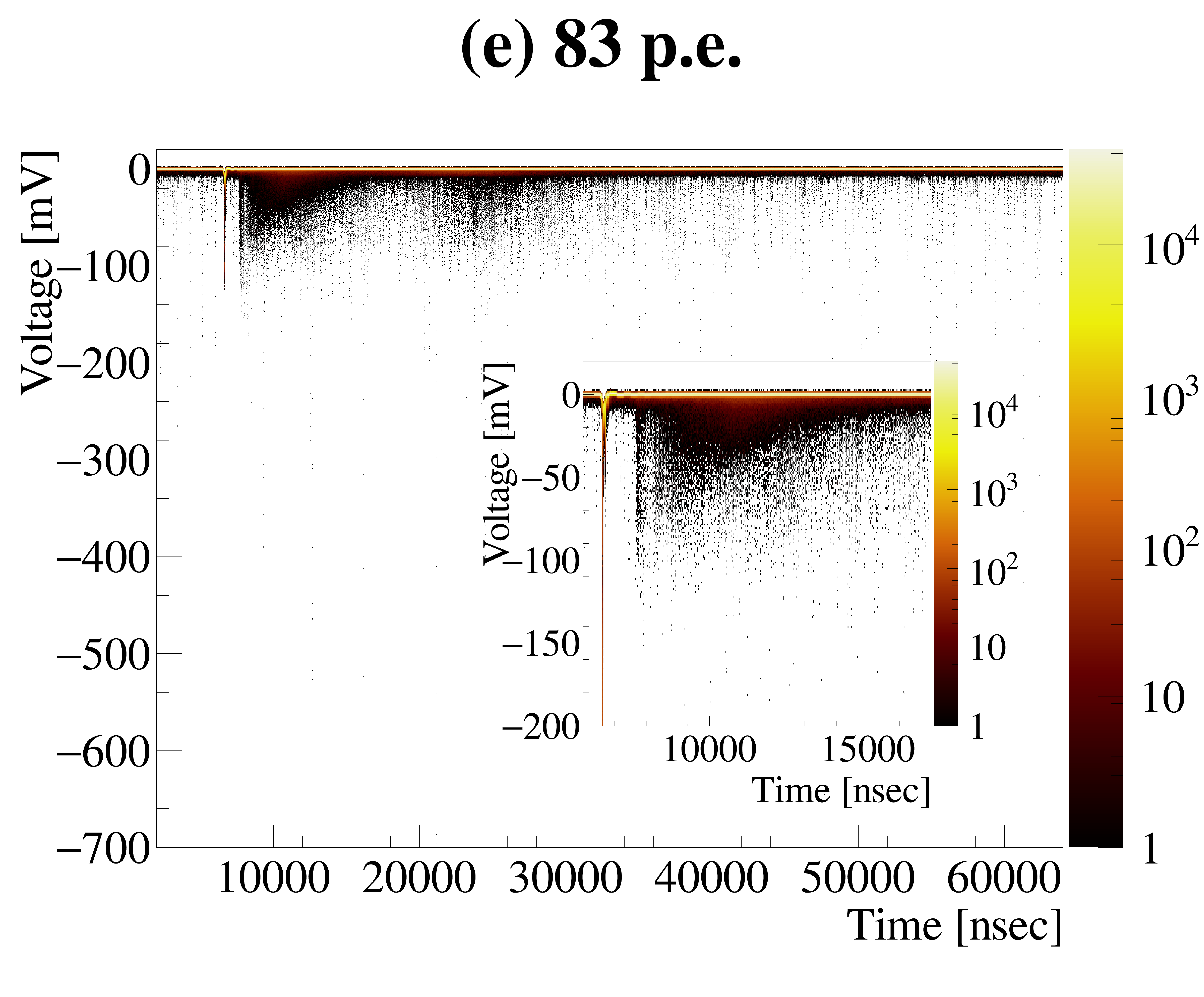}}
    \subfloat{\includegraphics[width=0.5\textwidth]{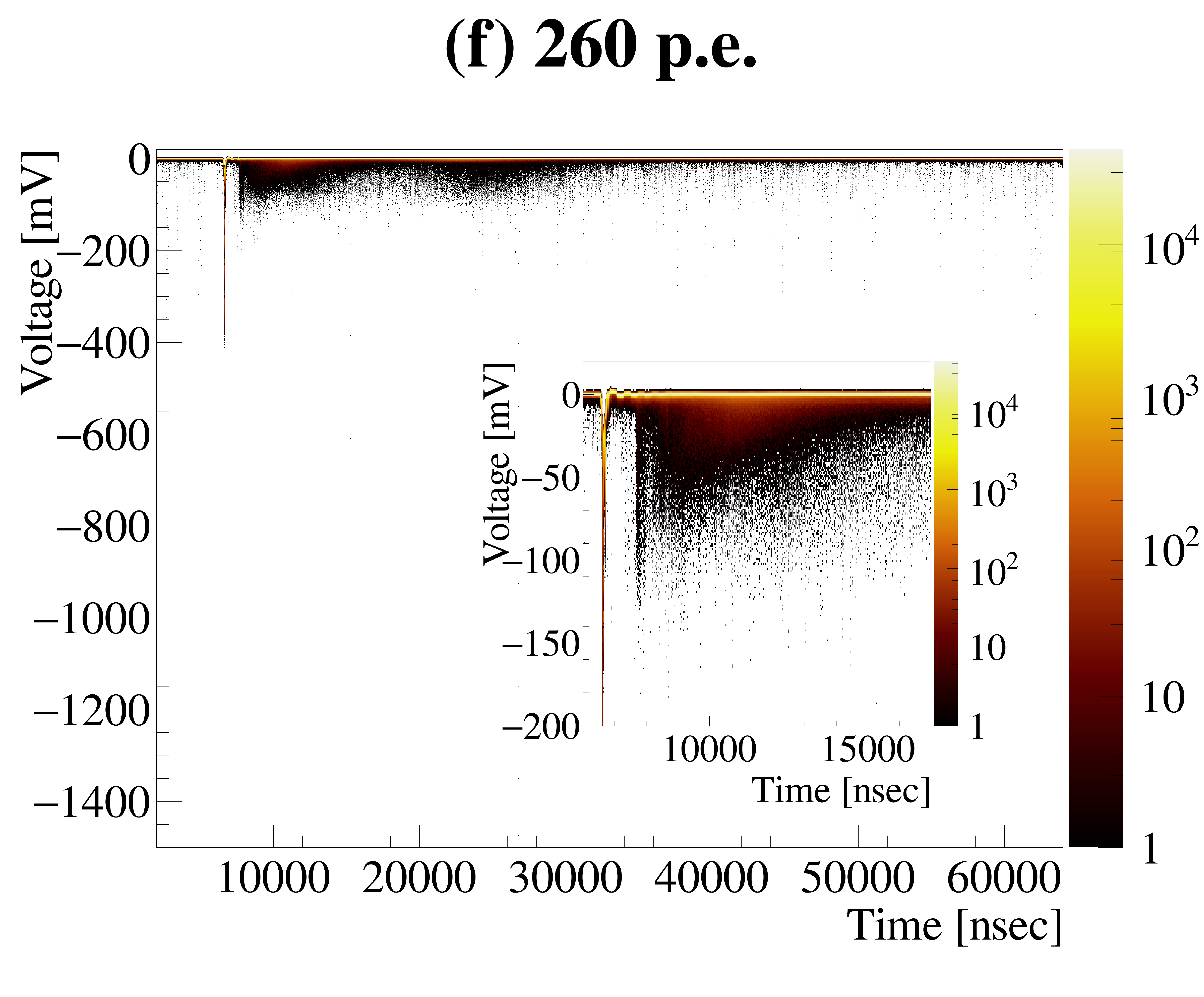}}
\caption{Accumulated pulses for 60 $\mu$sec at various light intensities for PMT A. The inset figures present zoomed-in views of the afterpulses with red dotted lines indicating the 150 mV level.}
\label{fig:afterpulse_pmtA}    
\end{figure}

\begin{figure}[!]
    \centering
    \subfloat{\includegraphics[width=0.5\textwidth]{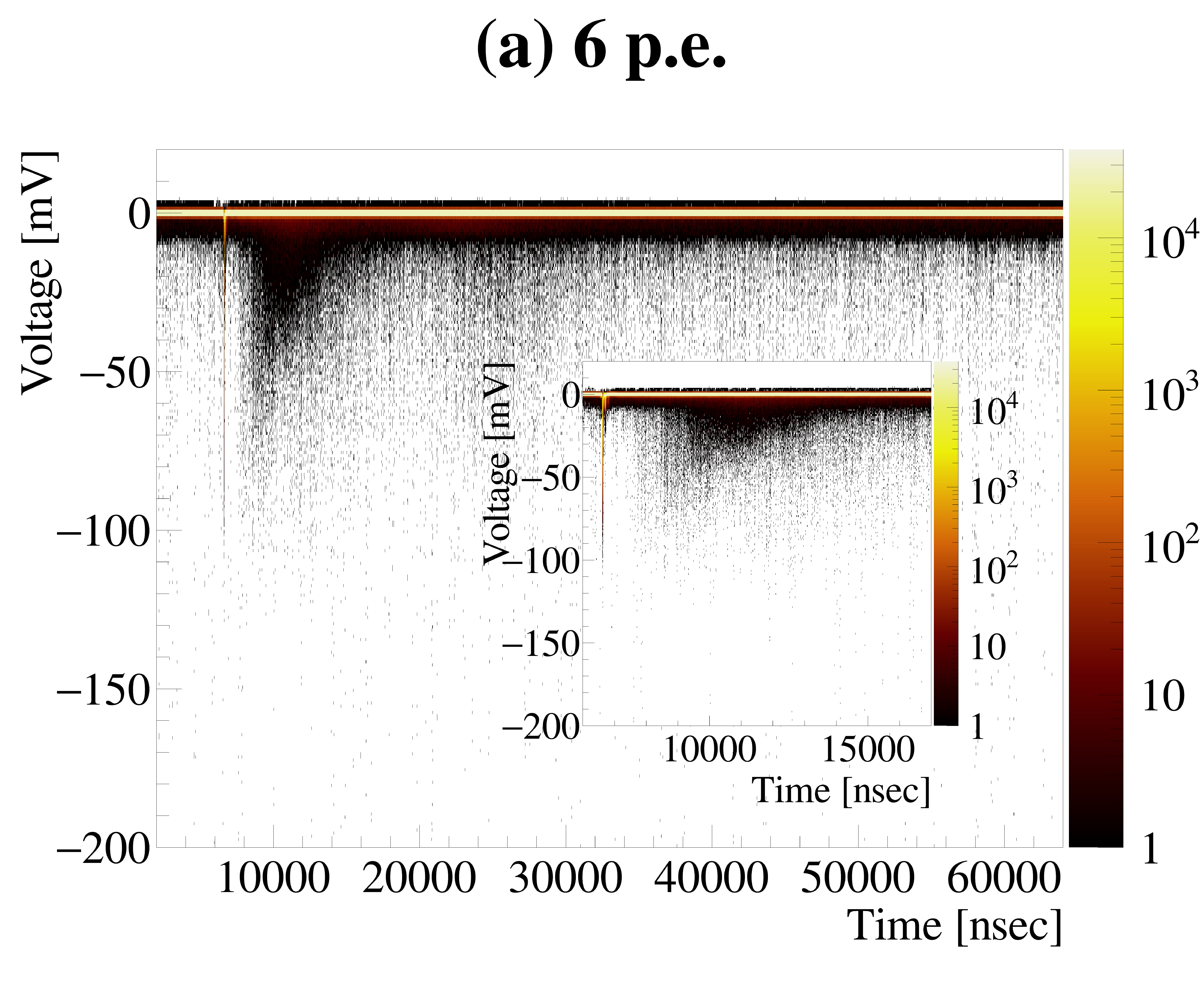}}
    \subfloat{\includegraphics[width=0.5\textwidth]{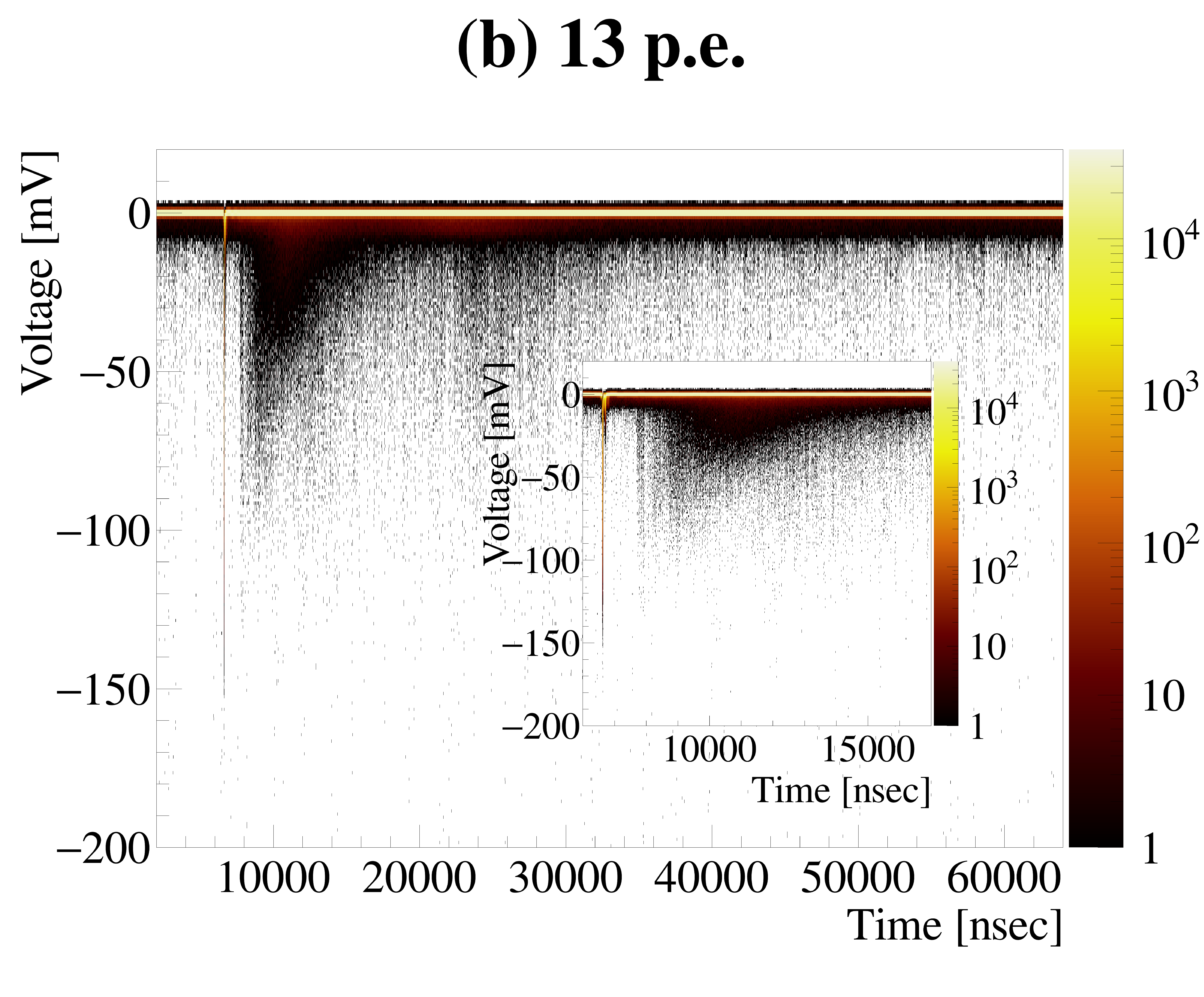}}\\
    \subfloat{\includegraphics[width=0.5\textwidth]{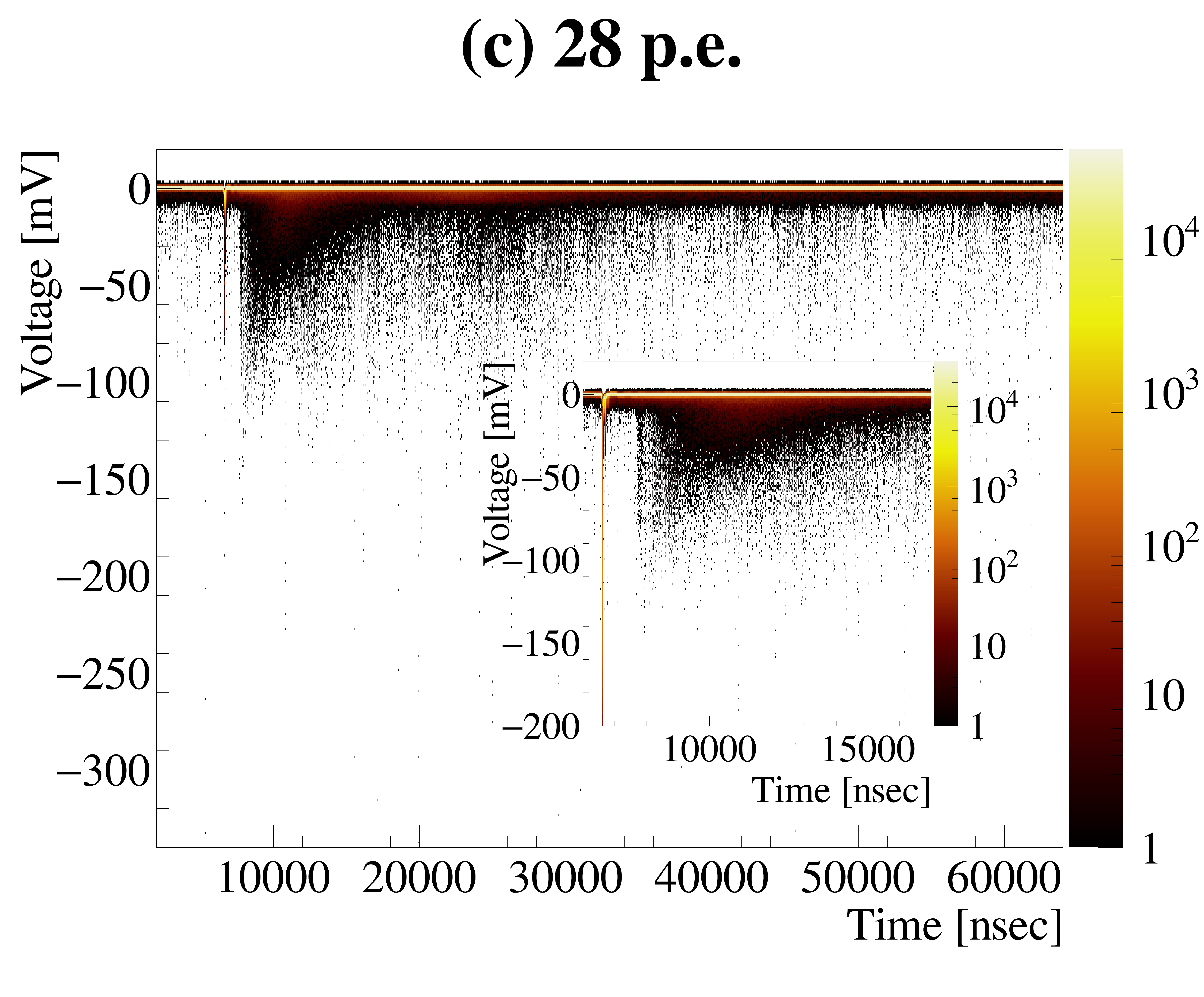}}   
    \subfloat{\includegraphics[width=0.5\textwidth]{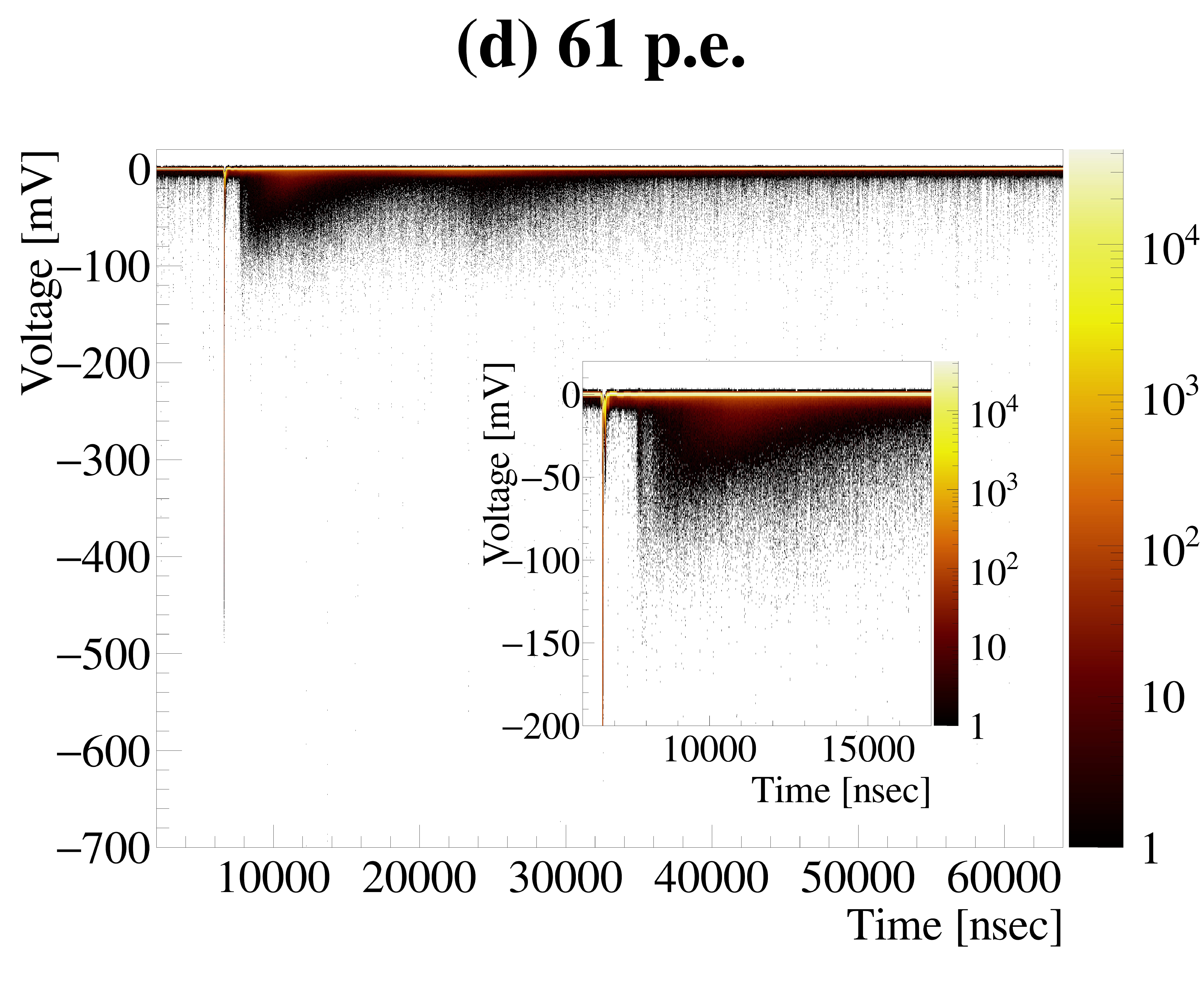}}\\
    \subfloat{\includegraphics[width=0.5\textwidth]{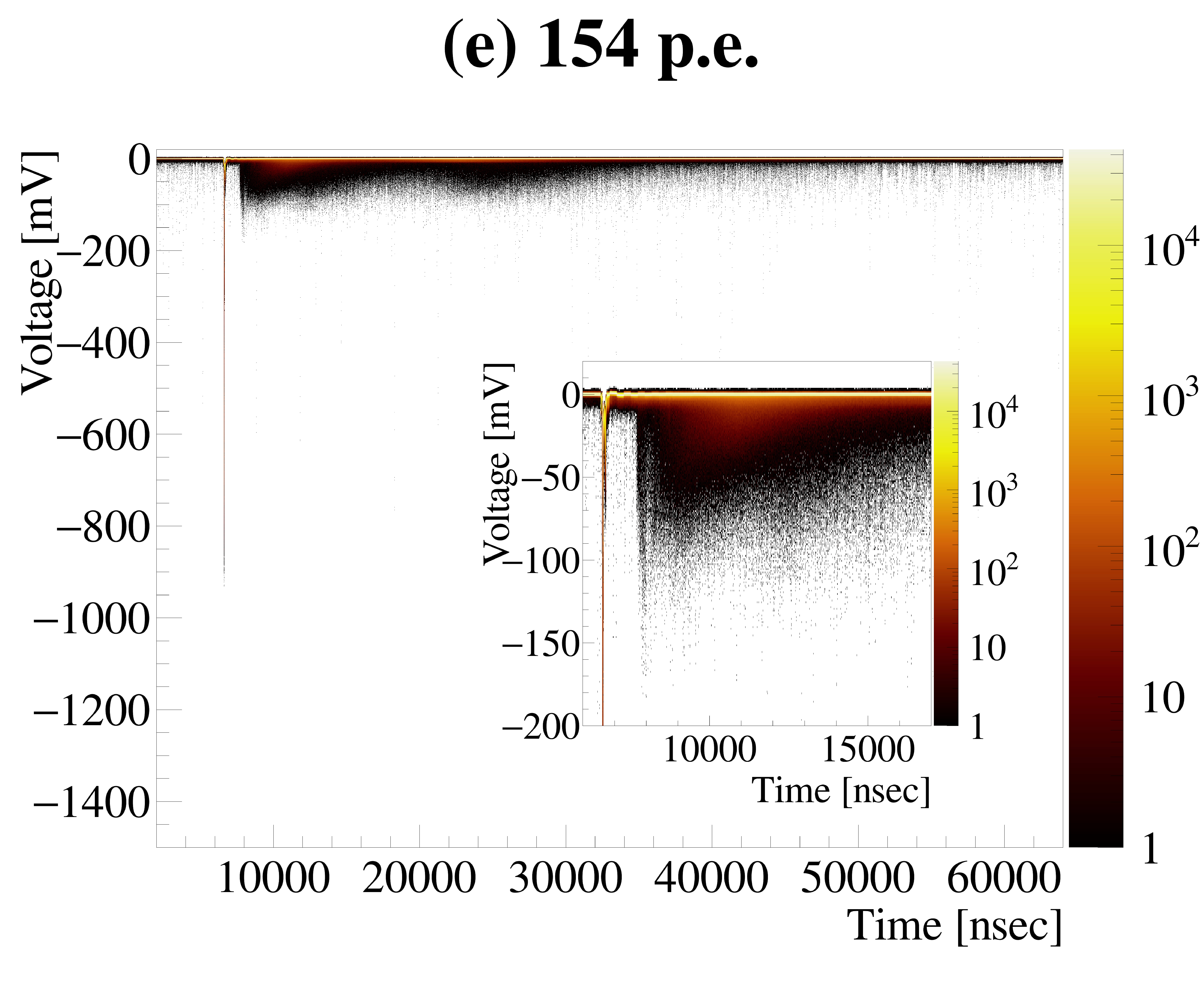}}
    \subfloat{\includegraphics[width=0.5\textwidth]{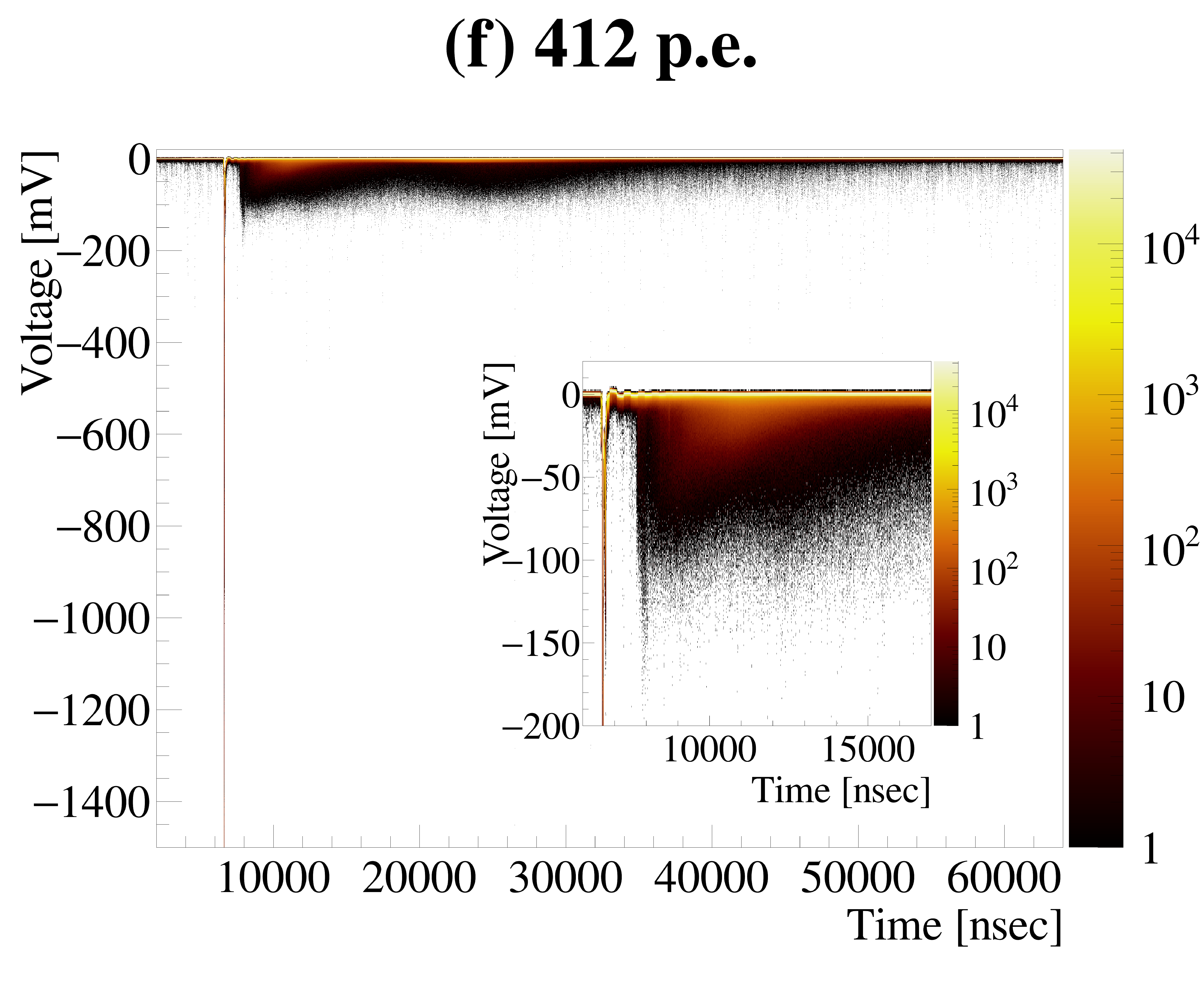}}
\caption{Accumulated pulses for 60 $\mu$sec by the various light intensities for PMT B. The inset panels provide zoomed-in views of the afterpulses with red dotted lines indicating the 150 mV level.}
\label{fig:afterpulse_pmtB}
\end{figure}

\begin{figure}[!]
\centering
\includegraphics[width=0.95 \textwidth]{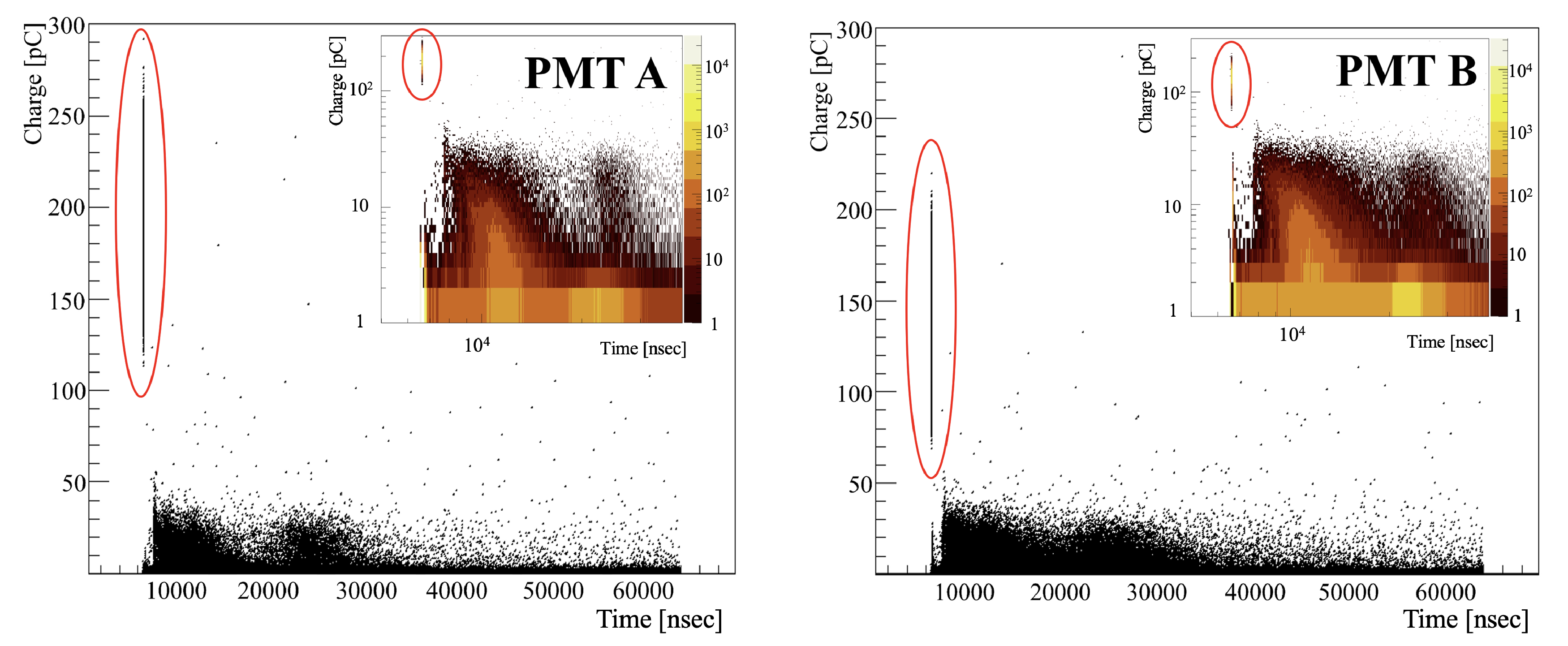}
    \caption{Representative plots of the charge versus time for all pulses recorded during 60 $\mu$sec~\cite{RENE}. Events within the red circle correspond to the main pulse. The inset plots use a logarithmic scale to enhance visibility. The main charges in the left and right plots are approximately 190 pC and 145 pC respectively at a PMT gain of $1\times10^{7}$, corresponding to $\sim$120 p.e. and $\sim$90 p.e.}
    \label{fig:time_afterpulse}
\end{figure}

Afterpulse are caused by the ionization of the residual gases within the PMT~\cite{afterpulse1, afterpulse2} and occur several hundred nanoseconds to microseconds after the main pulse. Notably, the RENE experiment detects the neutrinos via the inverse beta decay ($\bar{\nu_e}+ p\rightarrow e^{+} + n$) process. To select neutrino events, the prompt signal from the positron and the delayed signal from the neutron are paired based on a timing constraint. However, afterpulses can mimic either the prompt or delayed signals, and contaminate the neutrino dataset. Consequently, characterizing their behavior is essential. Fig.~\ref{fig:afterpulse_pmtA} and ~\ref{fig:afterpulse_pmtB} display accumulated pulses for each PMT, generated from 50,000 events for 60 $\mu$sec at various light intensities. Afterpulses appear at consistent timing regardless of light intensity. As the light intensity increases, the number of afterpulses increases, while the pulse height---which reflects the charge---reaches a saturation point. The charges as a function of time are plotted in Fig.~\ref{fig:time_afterpulse}. Especially, PMT B, despite being operated at a lower light intensity, exhibits a higher afterpulse rate than PMT A, which was exposed to a higher light intensity. This means the afterpulse rate differs for each PMT, which indicates that gas impurities might vary from one PMT to another.

\begin{figure} [!htb]
\centering
\includegraphics[width=1.0 \textwidth]{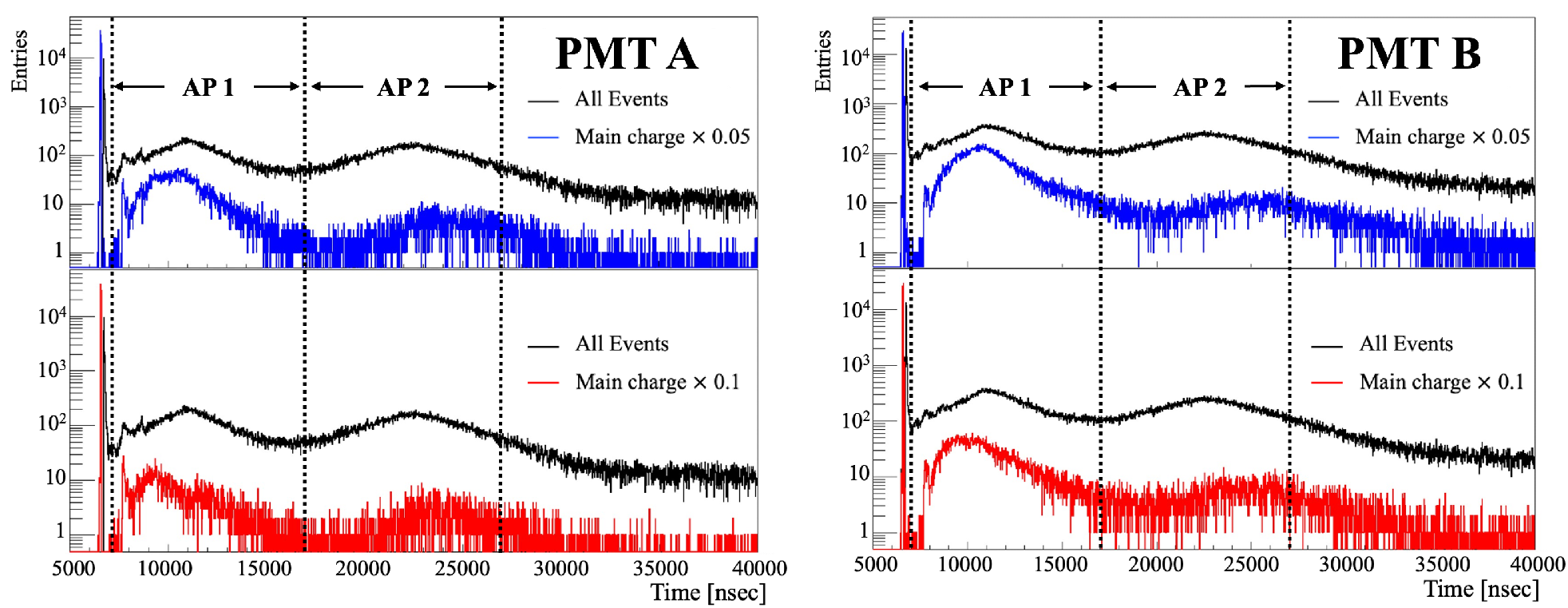}
  \caption{Event rate as a function of event time for PMT A (left) and PMT B (right), corresponding to the pulses shown in Fig.~\ref{fig:time_afterpulse}~\cite{RENE}. The upper panels display all recorded events (black) and the selected events with charges greater than 0.05 times the main charge (blue). The lower panels show all events (black) and selected events with charges exceeding 0.1 times the main charge (red).}
  \label{fig:event rate of afterpulse}
\end{figure}

\begin{figure}[!]
    \centering
    \subfloat{\includegraphics[width=0.5\textwidth]{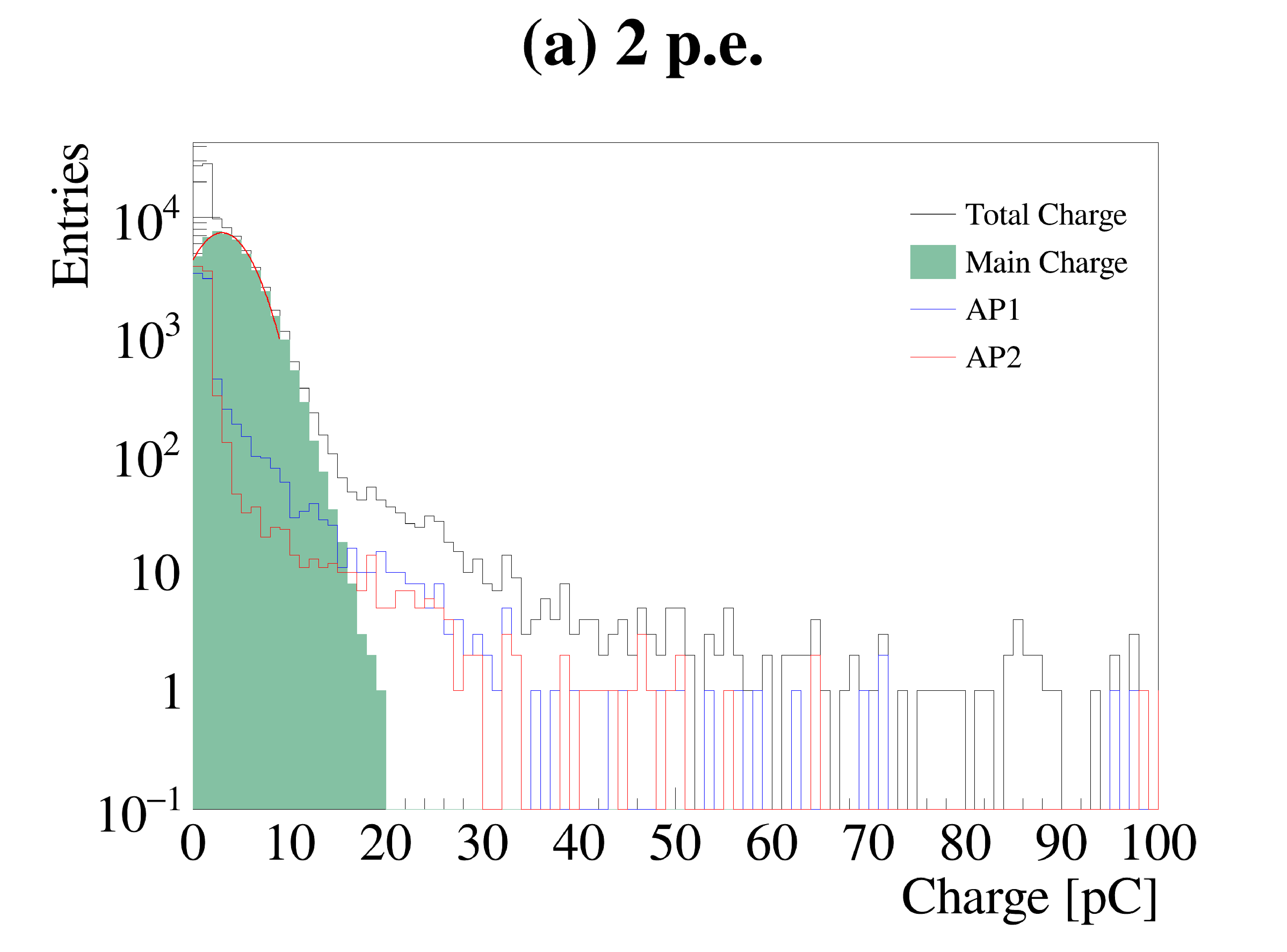}}
    \subfloat{\includegraphics[width=0.5\textwidth]{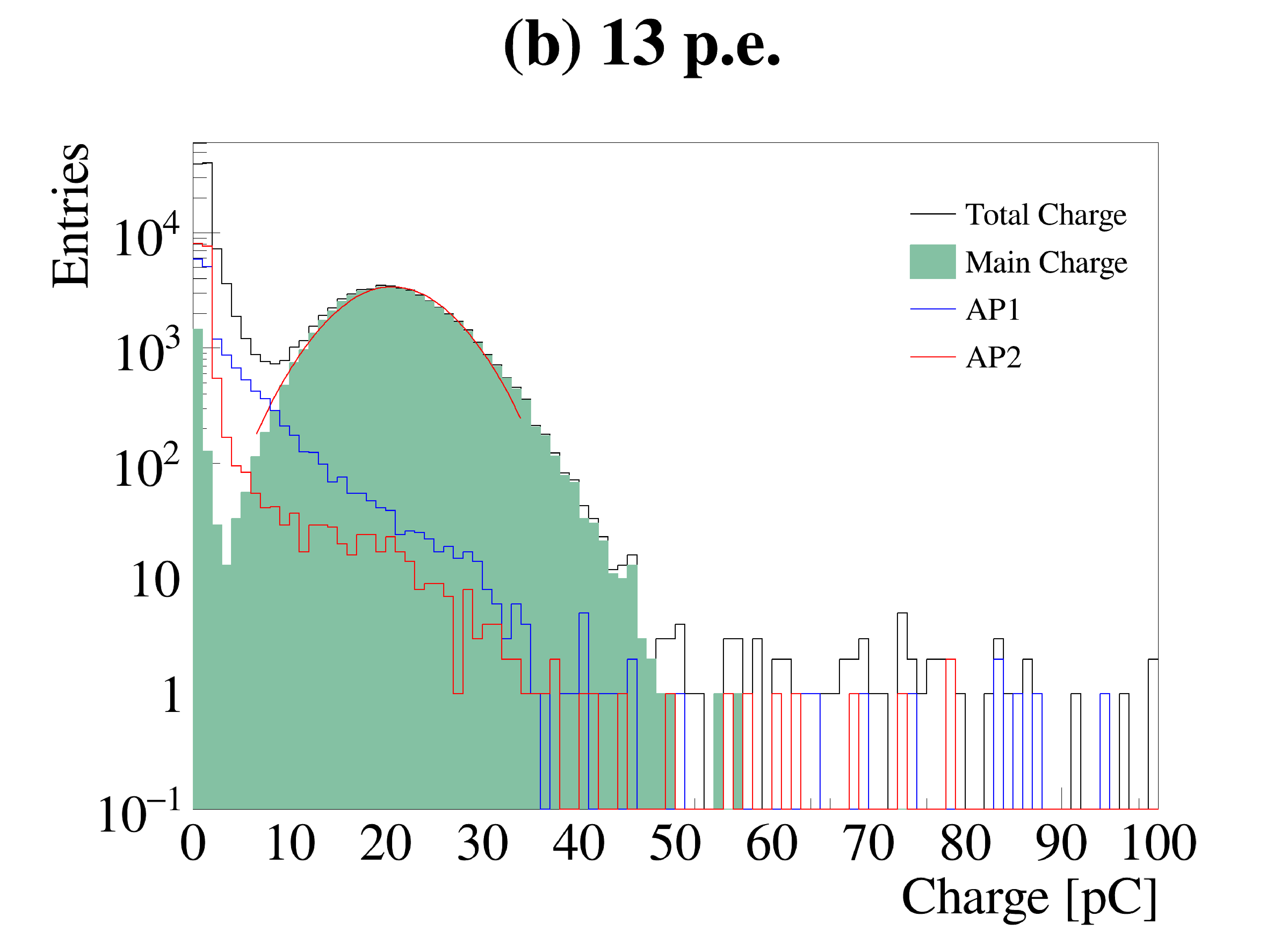}}\\
    \subfloat{\includegraphics[width=0.5\textwidth]{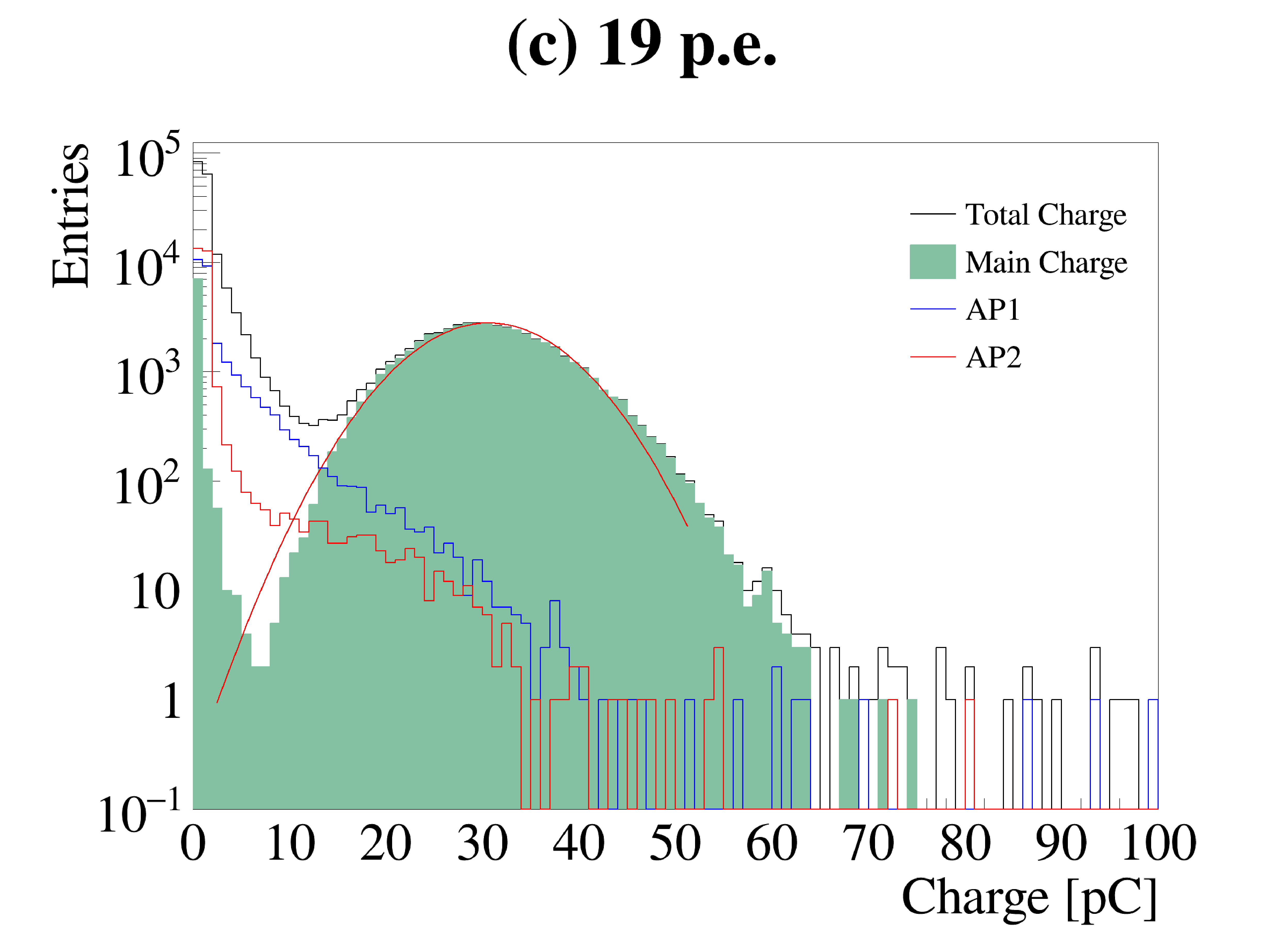}}
    \subfloat{\includegraphics[width=0.5\textwidth]{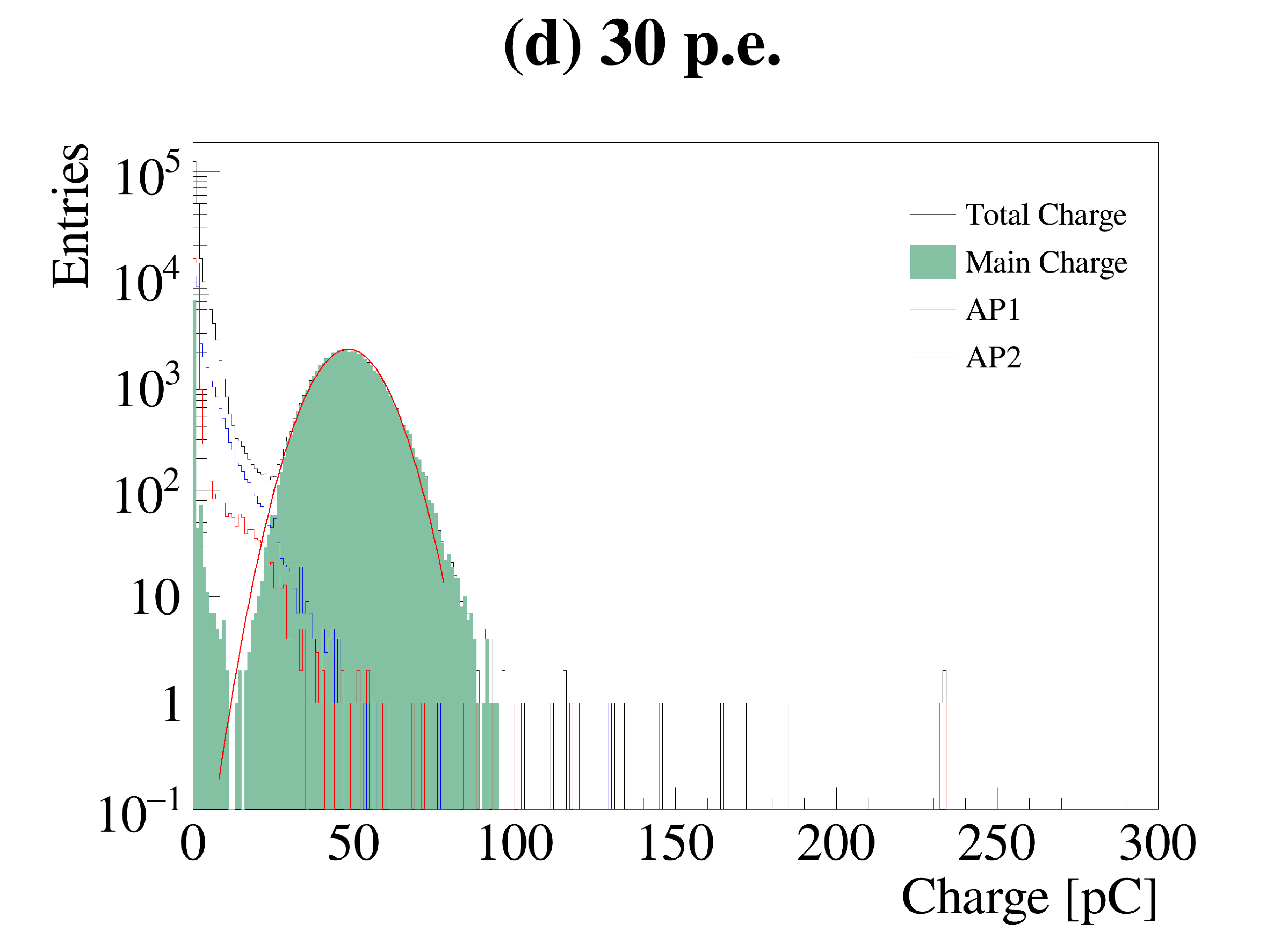}}\\
    \subfloat{\includegraphics[width=0.5\textwidth]{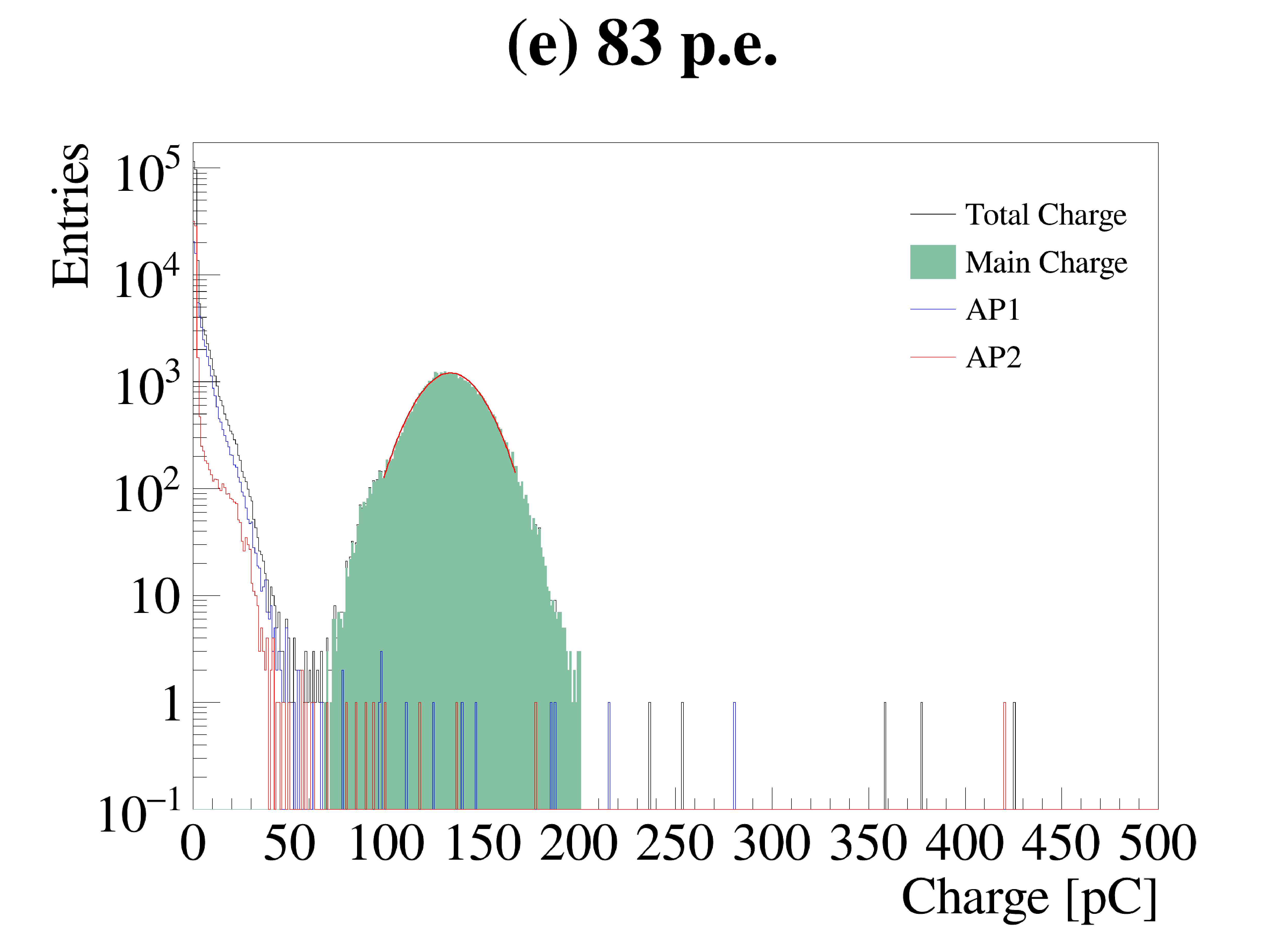}}
    \subfloat{\includegraphics[width=0.5\textwidth]{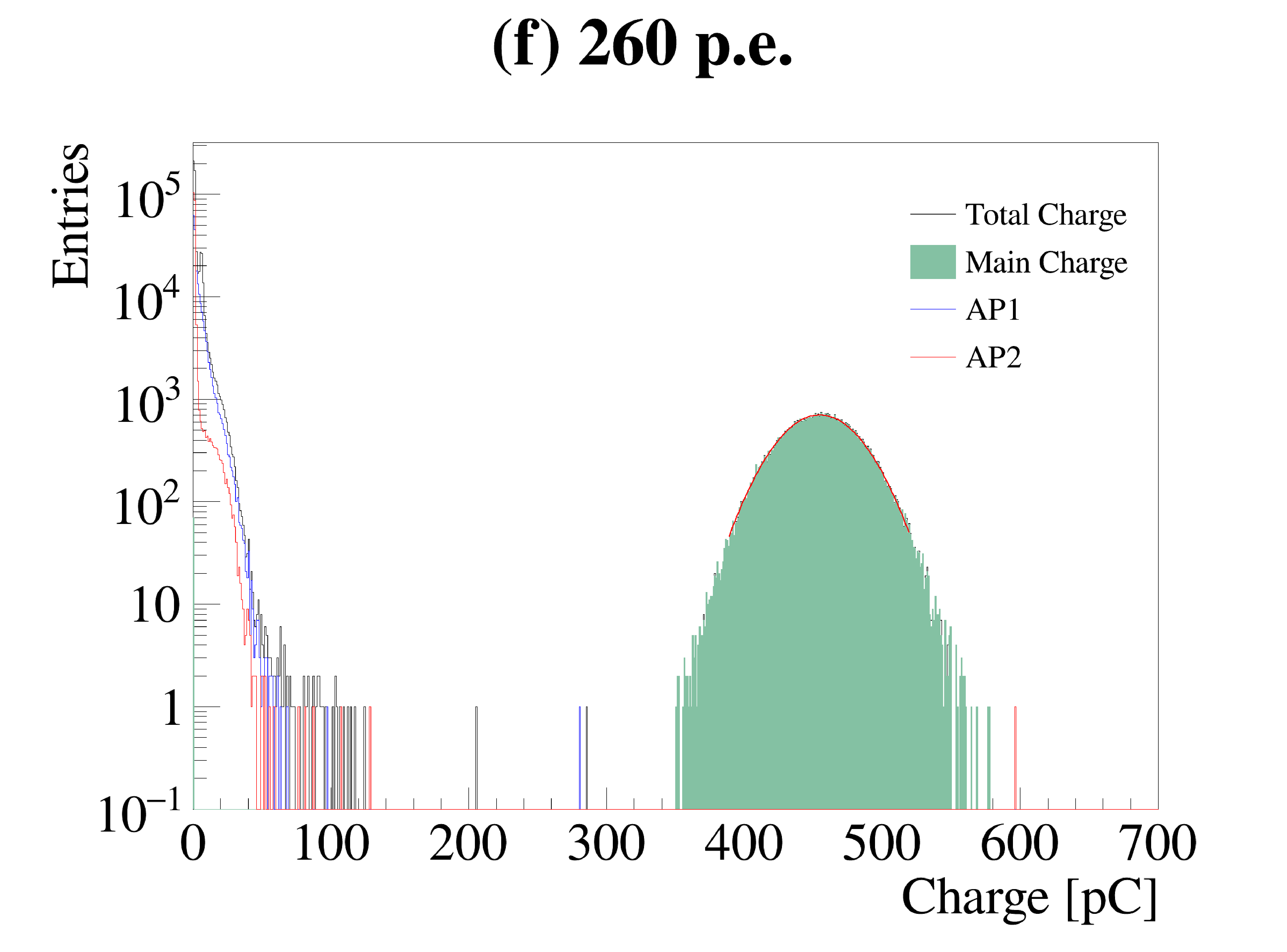}}
    \caption{Charge distributions of the afterpulses for PMT A.}
    \label{fig:charge_afterpulse_pmtA}
\end{figure}

\begin{figure}[!]
    \centering
    \subfloat{\includegraphics[width=0.5\textwidth]{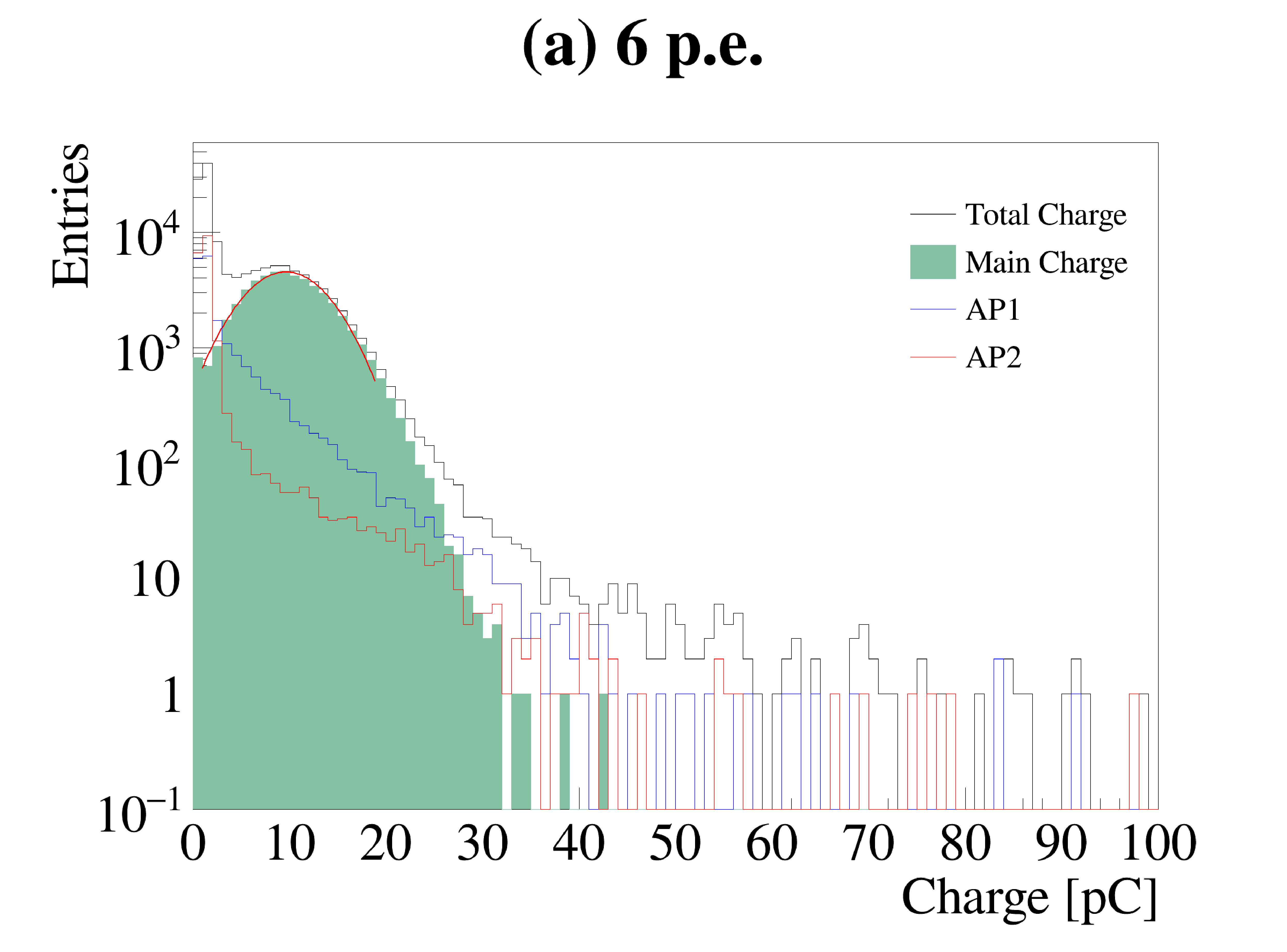}}
    \subfloat{\includegraphics[width=0.5\textwidth]{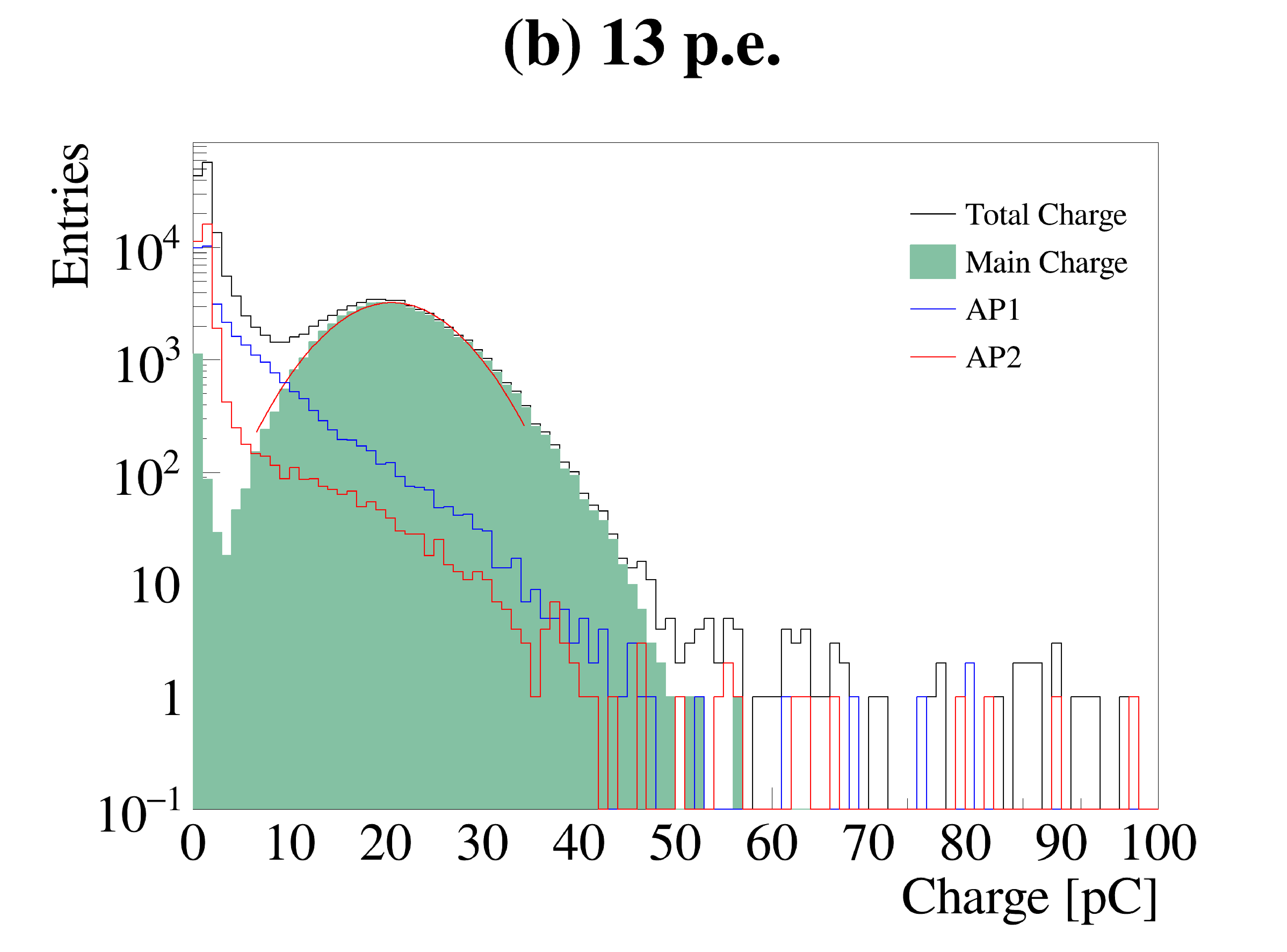}}\\
    \subfloat{\includegraphics[width=0.5\textwidth]{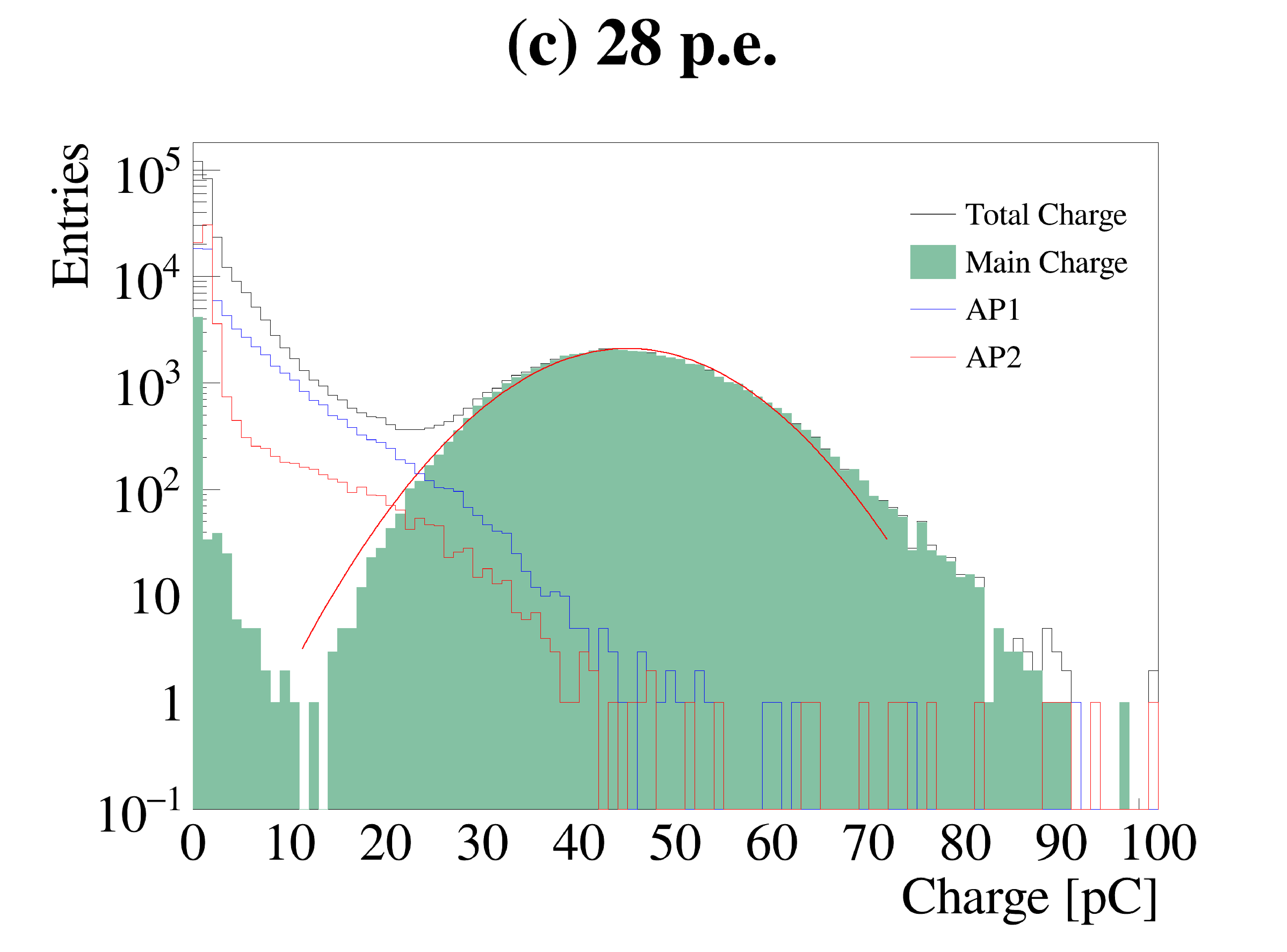}} 
    \subfloat{\includegraphics[width=0.5\textwidth]{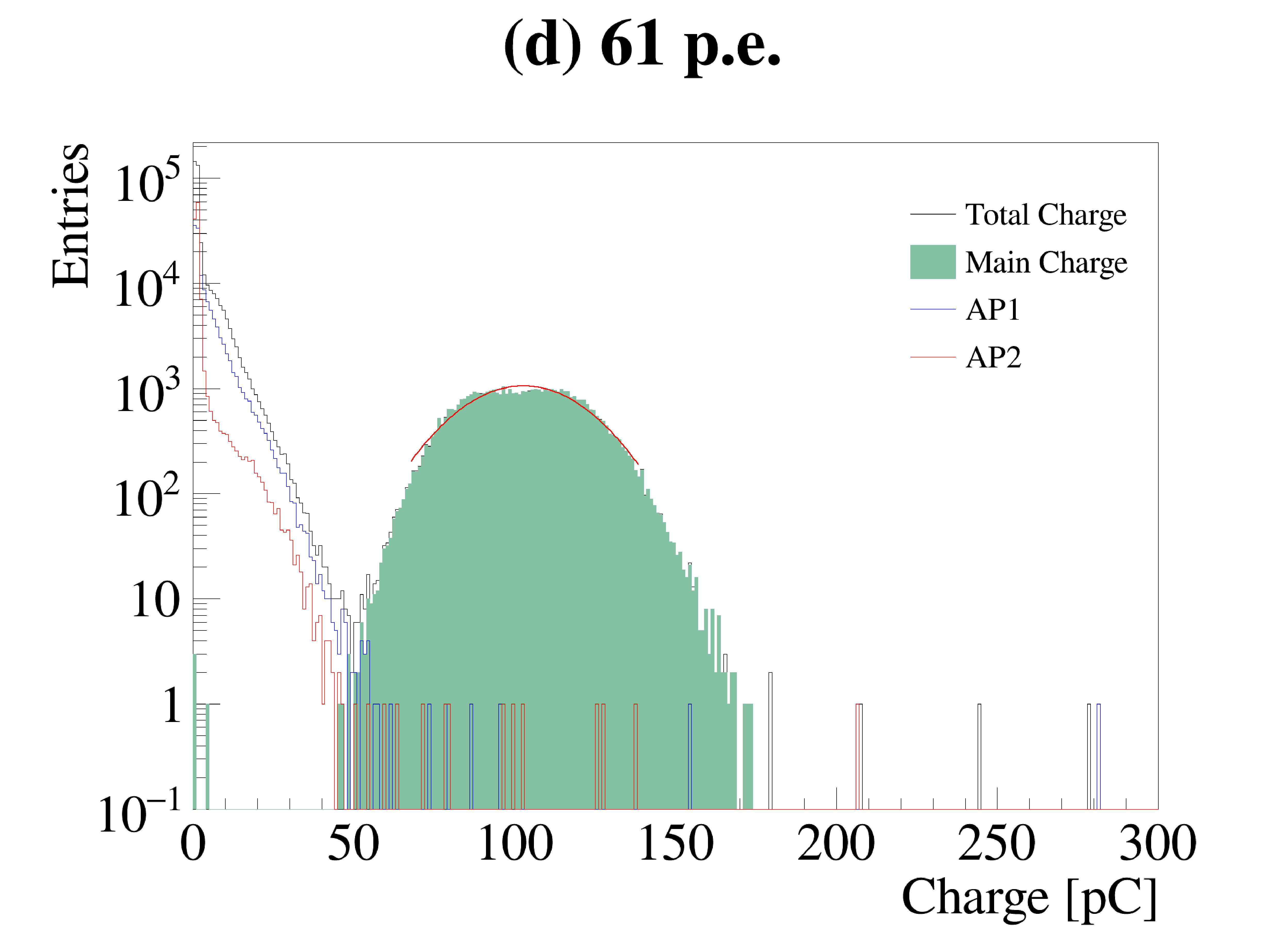}}\\
    \subfloat{\includegraphics[width=0.5\textwidth]{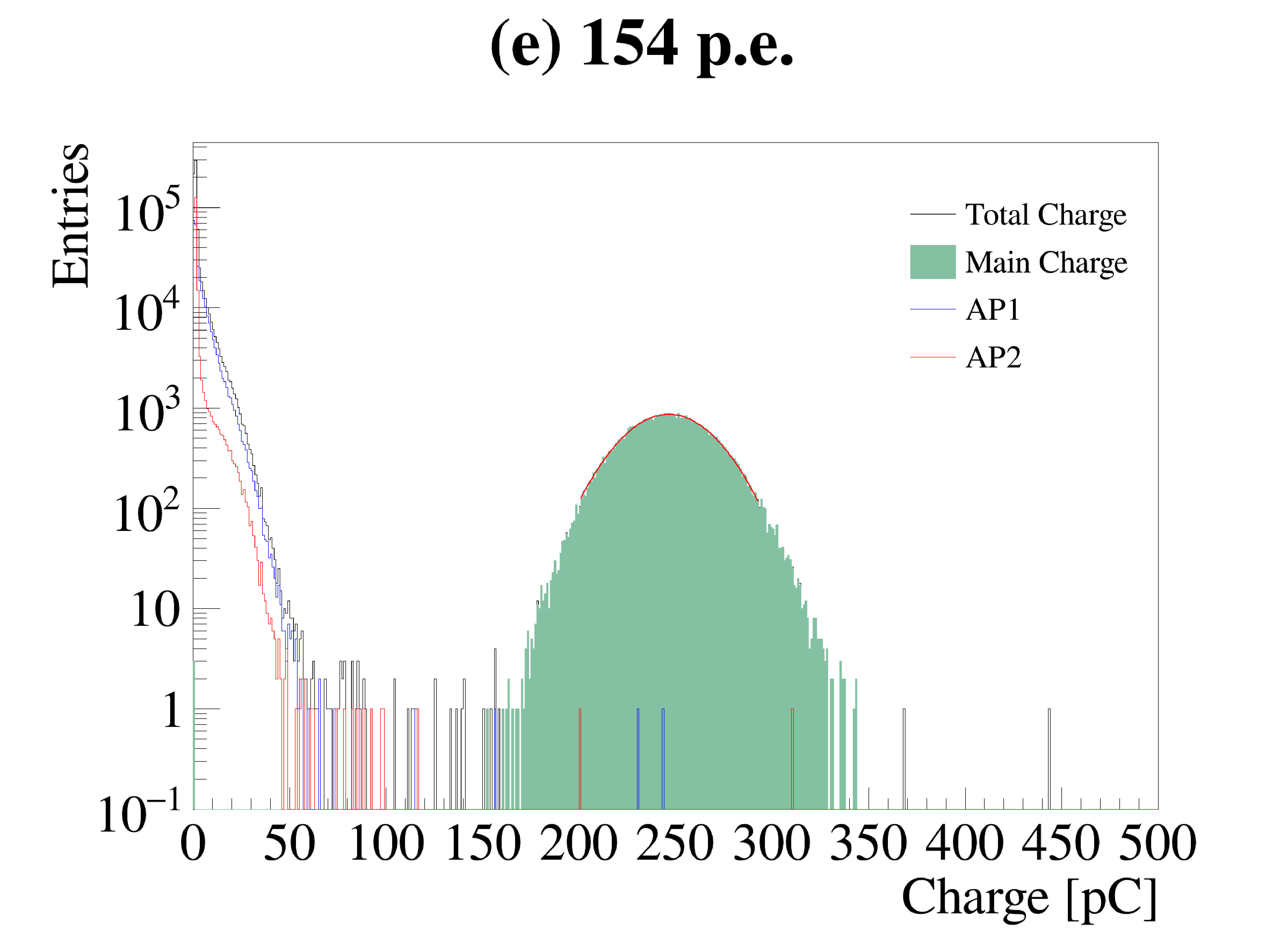}}
    \subfloat{\includegraphics[width=0.5\textwidth]{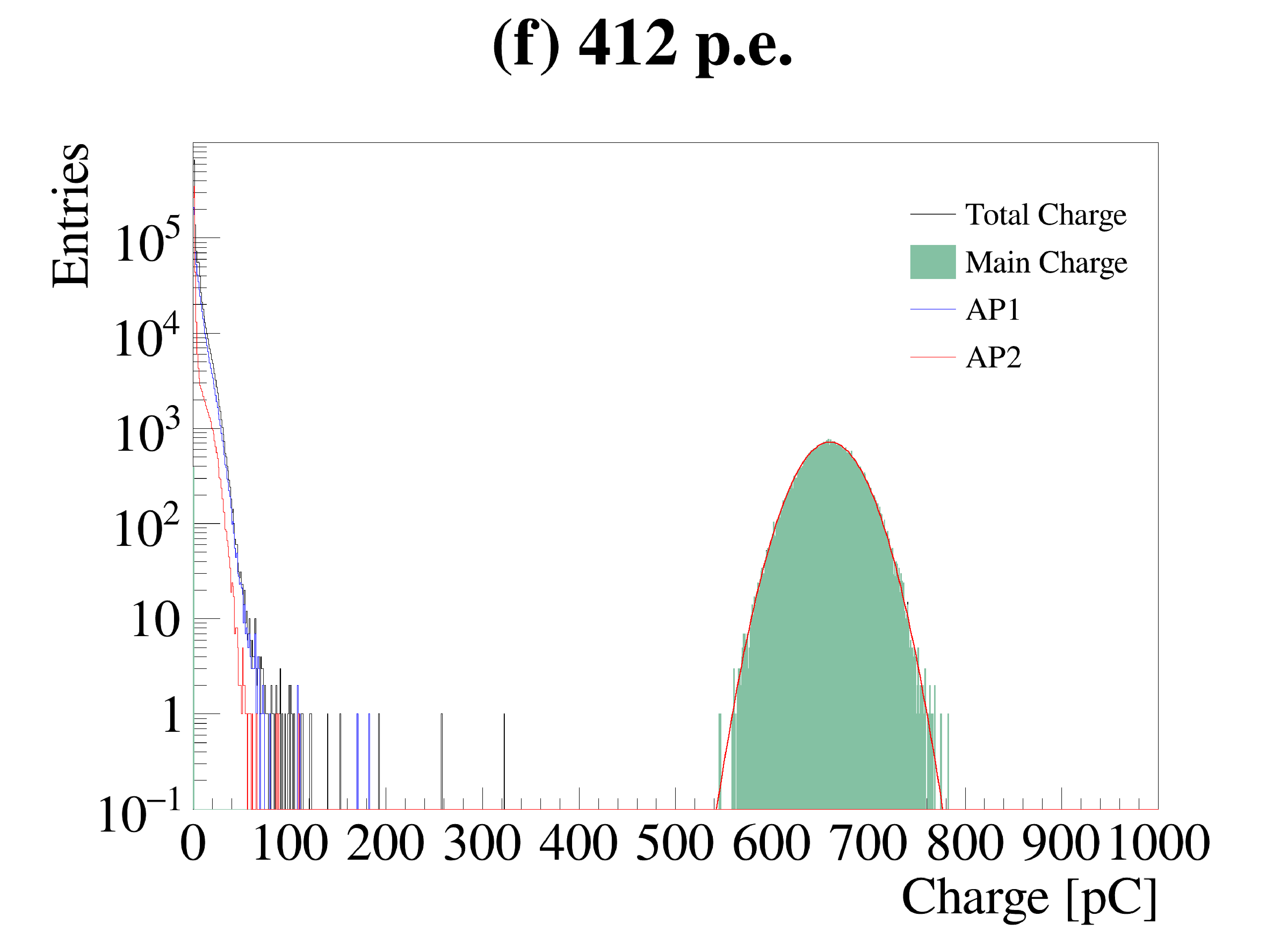}}
    \caption{Charge distributions of the afterpulses for PMT B.}
    \label{fig:charge_afterpulse_pmtB}
\end{figure}

\begin{figure} [!htb]
  \subfloat{\includegraphics[width=0.5\textwidth]{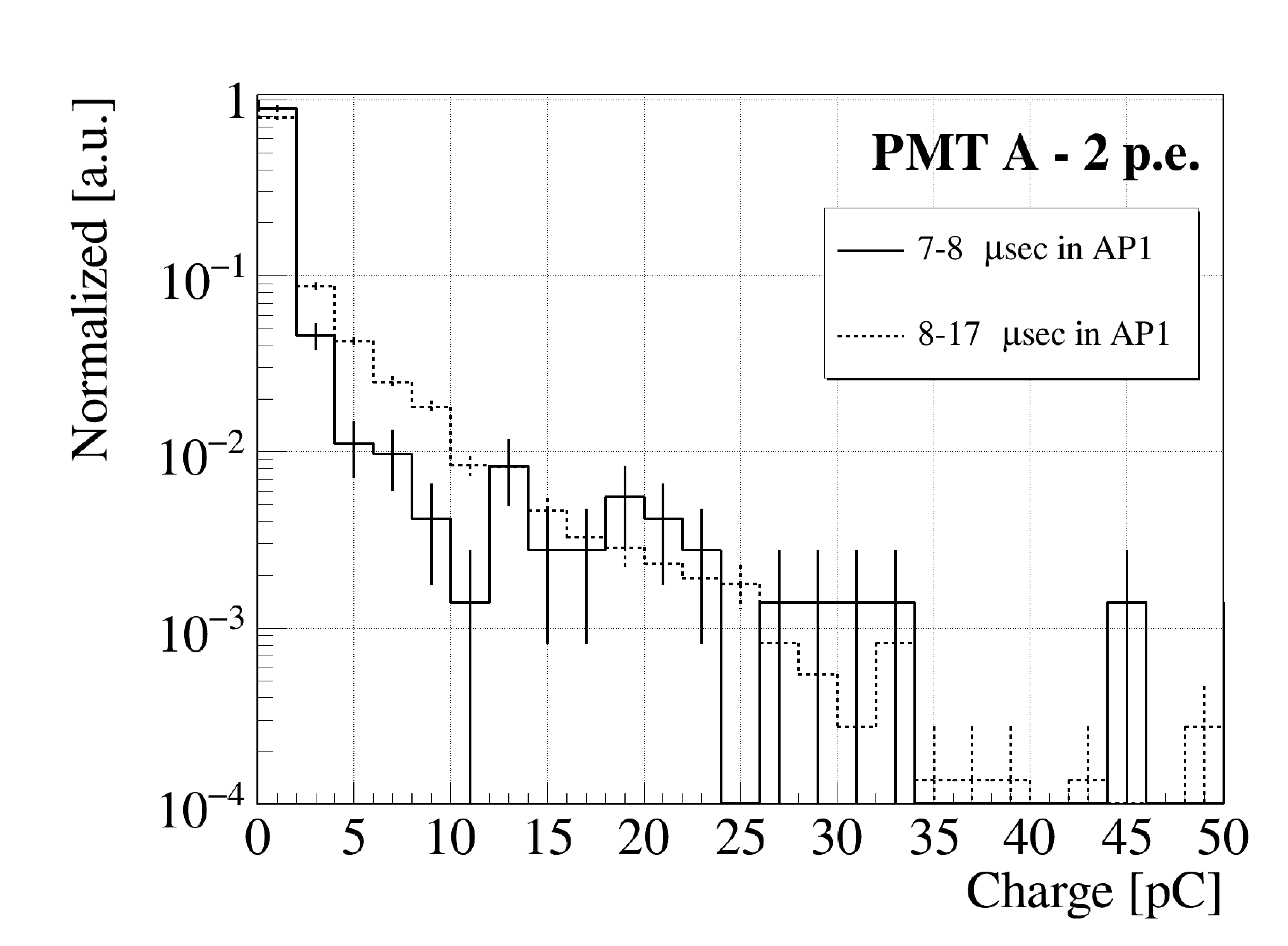}}
  \subfloat{\includegraphics[width=0.5\textwidth]{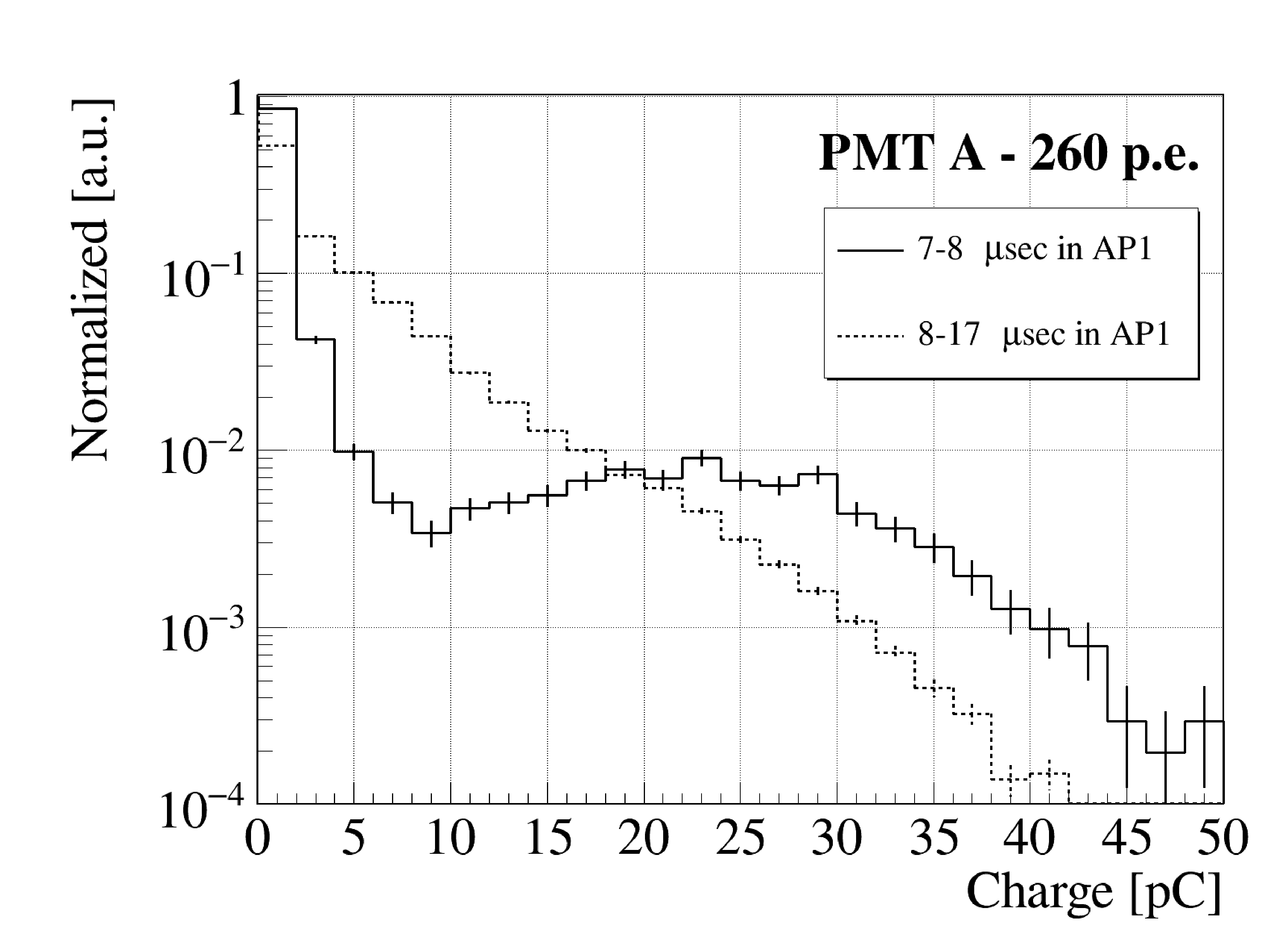}}
\caption{Charge distributions of the afterpulses in the AP1 region with a main charge of 2 p.e.(left) and 260 p.e.(right) for PMT A.}
\label{fig:AP1_PMTA}
\end{figure}

\begin{figure} [!htb]
  \subfloat{\includegraphics[width=0.5\textwidth]{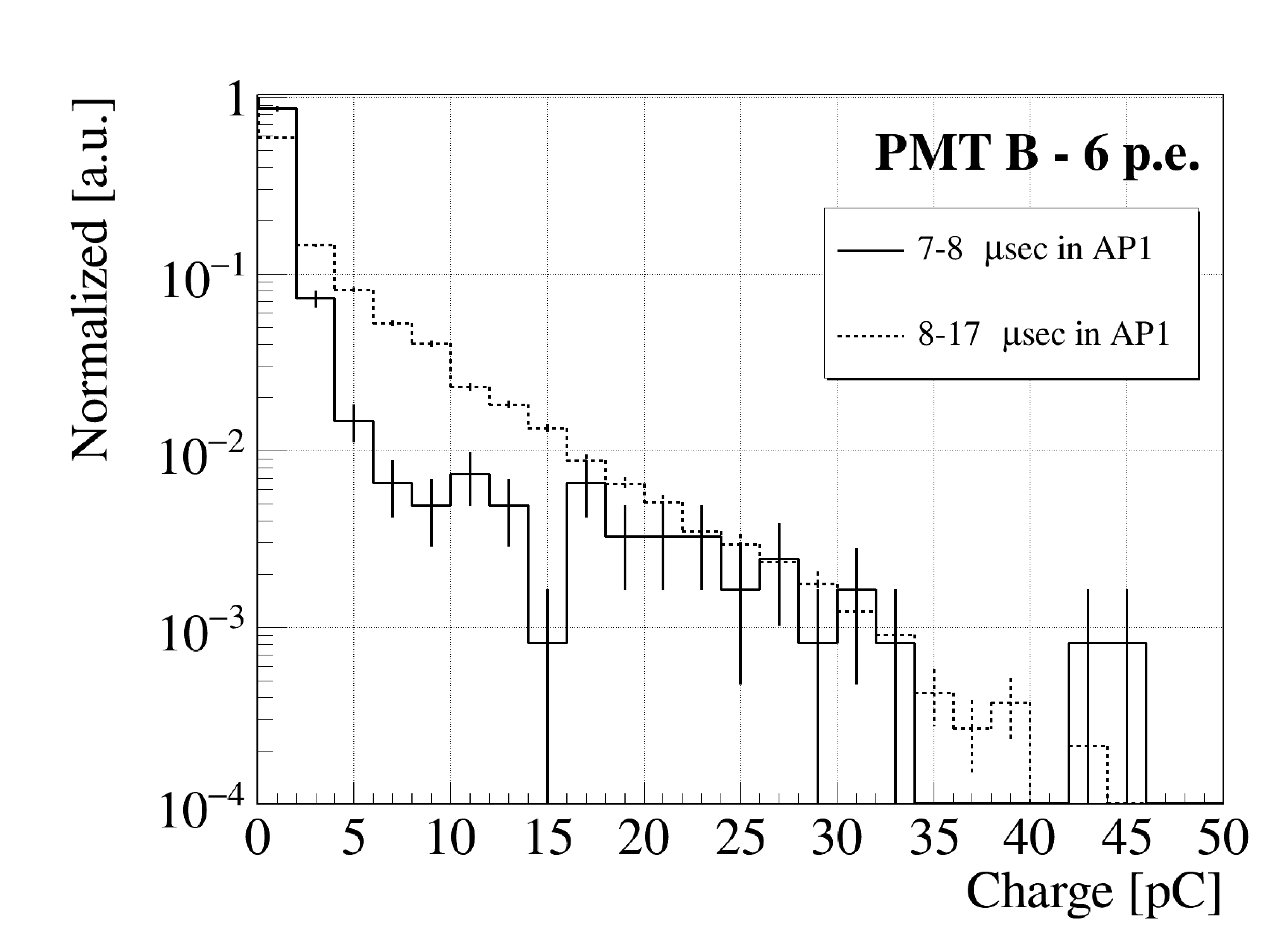}}
  \subfloat{\includegraphics[width=0.5\textwidth]{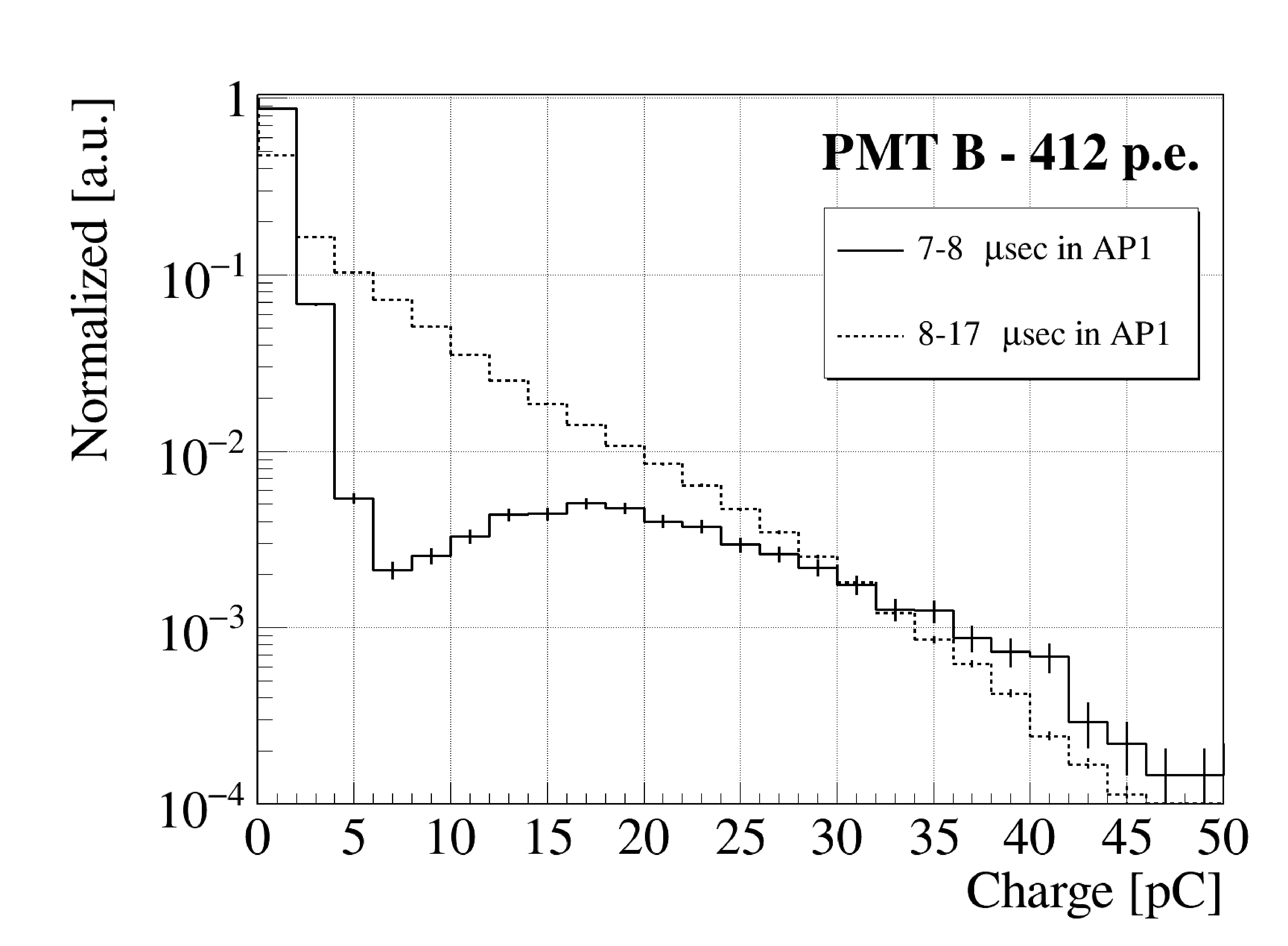}}
\caption{Charge distributions of the afterpulses in the AP1 region with a main charge of 6 p.e.(left) and 412 p.e.(right) for PMT B.}
\label{fig:AP1_PMTB}
\end{figure}

While the occurrence rate of afterpulses differs between the PMTs, the timing of their appearance is consistent across both PMTs, as shown in Fig.~\ref{fig:event rate of afterpulse}. In particular, the afterpulse timing can be categorized into two 10 $\mu$sec intervals starting 500 nsec after main pulse. The first 10 $\mu$sec (7--17 $\mu$sec) was defined as the AP1 region and the second 10 $\mu$sec (17--27 $\mu$sec) was labeled the AP2 region. The upper panels of Fig.~\ref{fig:event rate of afterpulse} display the rate of afterpulses with charge exceeding $5\%$ of the main pulse charge, while the lower panels present the rate of those exceeding $10\%$. In the second 10 $\mu$sec time intervals, the event rate of afterpulses clearly decreases compared to the first 10 $\mu$sec time intervals. Based on the timing distribution of afterpulses shown in Fig.~\ref{fig:event rate of afterpulse}, the afterpulse charge was evaluated for each time interval. Fig.~\ref{fig:charge_afterpulse_pmtA} and ~\ref{fig:charge_afterpulse_pmtB} present the charge distributions for each timing interval. The green-colored distributions represent the charges of the main pulses, while the blue and red distributions correspond to the afterpulse charges in the AP1 and AP2 regions, respectively. The afterpulse charge gives a clear upper limit, with the maximum observed value reaching approximately 30 p.e. for both 20-inch PMTs across all measurements. This upper bound is also evident in Fig.~\ref{fig:afterpulse_pmtA}, ~\ref{fig:afterpulse_pmtB} and ~\ref{fig:time_afterpulse} and consistent with previous findings~\cite{semuli, afterpulse2, afterpulse3}. The afterpulse in a photomultiplier tube originates from secondary electrons emitted when positive ions, produced from residual gas inside of PMT, are accelerated by the internal electric field and subsequently strike the photocathode. The charge of the afterpulse can vary depending on the type of ion and the location at which the ion is generated, since these factors determine the ion’s drift path and the energy it acquires before reaching the photocathode. Nevertheless, the energy gained by a positive ion is fundamentally limited by the voltage distribution inside the PMT, typically on the order of a few hundred electronvolts. As a result, the number of secondary electrons that can be released from the photocathode is inherently bounded, leading to an upper limit on the afterpulse charge that is largely independent of the main pulse height. Fig.~\ref{fig:AP1_PMTA} and ~\ref{fig:AP1_PMTB} show the charge distributions of afterpulses occurring in two different time windows (7--8 $\mu$sec and 8--17 $\mu$sec) within the AP1 region shown in Fig.~\ref{fig:event rate of afterpulse}. The distinct peak observed at 7--8 $\mu$sec originate from ions with very short drift times, corresponding to afterpulses occurring within approximately 1.5 $\mu$sec after the main pulse observed at around 6.5 $\mu$sec. These afterpulses are predominantly concentrated in the low-charge region under low light intensity corresponding to a few p.e. However, as the incident light intensity increases, the charge distribution shifts noticeably toward higher values, as observed in both PMTs as shown in Fig.~\ref{fig:AP1_PMTA} and ~\ref{fig:AP1_PMTB}. This behavior can be explained by the increased production of positive ions resulting from collisions between primary photoelectrons and residual gas molecules at higher light intensities. In particular, lighter ions with short drift times can arrive at the photocathode within a narrow time interval, and the resulting afterpulses may temporally overlap, leading to larger apparent single pulses~\cite{afterpulse4}. In contrast, afterpulses observed in the 8--17 $\mu$sec window shown in Fig.~\ref{fig:event rate of afterpulse}, corresponding to delay times greater than approximately 1.5 $\mu$sec after the main pulse, do not exhibit a significant change in their charge distribution with varying light intensity. These results indicate that afterpulses occurring in different time windows within the AP1 region exhibit distinct charge behaviors, suggesting that they may originate from different ion species or generation mechanisms. For the events occurring for the first 1 $\mu$sec in AP1 region, high-intensity signals exceeding 30 p.e. are occasionally observed; however, their occurrence rate is very low, accounting for less than $0.1\%$ of the main events. The RENE detector will rely on these two PMTs for reactor neutrino detection, with each PMT mostly receiving several hundred photoelectrons per event. Given the well-characterized timing of afterpulses and the observed upper limit of their charges, events from afterpulses are expected to be effectively excluded.

\section{Conclusion}

RENE experiment is prepared to investigate the origin of the reactor neutrino anomaly. The RENE detector is a Gd-loaded LS detector equipped with two 20-inch Hamamatsu R12860 PMTs. Prior to detector construction, the PMTs---serving as critical components---were evaluated under environmental conditions that closely implemented \textit{in-situ} conditions including temperature, humidity and magnetic field. The characteristic of the charge and timing response for the two PMTs were confirmed. In particular, the gain variation was examined as a function of the light injection position on the PMT cathode, since the entire 20-inch cathode surface is expected to contribute actively given the detector design. Based on the dynode structure, we scanned the axis expected to exhibit the most extreme gain variation across the entire cathode surface and confirmed that the gain varies by up to $\pm10\%$. The RENE detector is intended to operate at gains lower than the nominal $1 \times 10^7$ to prevent saturation of the PMTs or the DAQ system. Accordingly, gain stability and timing response were measured at a gain of $7\times10^6$. The gain remained within $\pm2\%$ over a period of 3,000 minutes, and the TTS was measured to be less than 4 nsec for both PMTs. Both late pulses and afterpulses were observed. The late pulses appeared approximately 100 nsec after the main pulse, with an occurrence rate below $1\%$ of the main pulse rate. The timing of the afterpulses showed a consistent pattern in both PMTs. Afterpulses began approximately 500 nsec after the main pulse, and two distinct structures were observed in the time distributions, each spanning 10 $\mu$sec. Because afterpulses can mimic the neutrino events in RENE, careful consideration of rejection strategies is essential. According to our measurements, the pulse height of the afterpulses consistently remained below an upper limit of 30 p.e. across all incident light intensities in both RENE PMTs. This result supports a straightforward approach for eliminating mimicked events, as detected signals exceeding 100 p.e. in each RENE PMT are expected to correspond to meaningful data for neutrino detection. The 20-inch R12860 PMTs, developed by modifying their dynode structure to enhance collection efficiency, are widely used in neutrino experiments. In this study, the basic performance of these PMTs and their suitability for the RENE experiment were evaluated. This report is expected to be useful not only for the simulation and data analyses for the RENE detector but also for other experimental investigations.

\section*{Acknowledgment}

This work was supported by the National Research Foundation of Korea(NRF) grant funded by the Korea government(MSIT) (RS-2021-NR060129, RS-2022-NR069287, RS-2022-NR070836, RS-2023-00211807, RS-2024-00416839, RS-2024-00442775, RS-2025-24533596, RS-2025-25400847, RS-2025-23525600, RS-2025-16071941).


%

\vspace{0.2cm}
\noindent


\let\doi\relax








\end{document}